\documentclass[12pt,preprint2]{aastex}
 \def\ltsima{$\; \buildrel < \over \sim
\;$} \def\simlt{\lower.5ex\hbox{\ltsima}}            % < over MMM
\def\gtsima{$\; \buildrel > \over \sim \;$}
\def\simgt{\lower.5ex\hbox{\gtsima}}            % > over MMM
\def\kms{Km~s$^{-1}$}

 \def\oviii{{\sc
Oviii}}
 \def\ovii{{\sc Ovii}}

 \def\kms{km~s$^{-1}$}

 \def\lia{Ly $\alpha$}

\begin{document} %\input tcilatex

\title{\sc The ionized nuclear environment in NGC 985 as seen by
Chandra and BeppoSAX}
\author{Y. Krongold\altaffilmark{1}, F. Nicastro\altaffilmark{1},
M. Elvis\altaffilmark{1}, N.S. Brickhouse\altaffilmark{1},
S. Mathur\altaffilmark{2}, A. Zezas\altaffilmark{1}}
\altaffiltext{1}{Harvard-Smithsonian Center for Astrophysics. 60
Garden Street, Cambridge MA 02138}
\altaffiltext{2}{Department of
Astronomy, Ohio State University, 140 West 18th Avenue, Columbus,
OH 43210}
%
%\author{Draft 5, Feb, 9 2004}

\begin{abstract}

\begin{small}
We investigate the ionized environment of the Seyfert 1 galaxy NGC
985 with a new {\em Chandra}-HETGS observation and an archival BeppoSAX
observation. Both spectra exhibit
strong residuals to a single powerlaw model, indicating the presence
of an ionized absorber and a soft excess. A detailed model over
the {\em Chandra} data shows that the 0.6-8 keV intrinsic continuum can be
well represented by a powerlaw ($\Gamma\approx1.6$) plus a blackbody
component (kT=0.1 keV). Two
absorption components are clearly required to fit the
absorption features observed in the {\em Chandra} spectrum.  The
components have a difference of 29 in ionization parameter and 3 in
column density. The presence of the low ionization component is
evidenced by an Fe M-shell unresolved transition array (UTA) produced
by charge states {\sc vii} to {\sc xiii}. The high ionization phase
is required by the presence of broad absorption features arising from
several blends of Fe L-shell transitions (Fe {\sc xvii-xxii}). A third
highly ionized component might also be present, but the data does not
allow to constrain its properties. Though poorly constrained, the
outflow velocities of the
components ($581\pm
206$ km s$^{-1}$ for the high ionization phase and $197\pm
184$ km s$^{-1}$ for the low ionization one)
are consistent with each other,
%(within 1.2$\sigma$)
and with the outflow velocities of the absorption components
observed in the UV. In addition, the low ionization component
produces significant amounts of O {\sc vi}, N {\sc v}, and C {\sc
iv} which suggests that a single outflow produces the UV and X-ray
features. The broad band (0.1-100 keV) continuum in the BeppoSAX
data can be parameterized by a powerlaw ($\Gamma\approx1.4$), a
blackbody (kT=0.1 keV), and a high energy cutoff (E$_c\approx 70$
keV). An X-ray luminosity variation by a factor 2.3 is observed
between the BeppoSAX and {\em Chandra} observations (separated by
almost 3 years). Variability in the opacity of the absorbers is
detected in response to the continuum variation, but while the
colder component is consistent with a simple picture of
photoionization equilibrium, the ionization state of the hotter
component seems to increase while the continuum flux drops. The most
striking result in our analysis is that during both the {\em
Chandra} and the BeppoSAX observations, the two absorbing
components appear to have the same pressure. Thus, we
suggest that the absorption arises from a multi-phase wind. Such
scenario can explain the change in opacity of both absorption
components during the observations, but requires that a third
hotter component is pressure-confining the two phases.  Hence, our
analysis points to a 3-phase medium similar to the wind found in
NGC 3783, and further suggests that such a wind might be a common
characteristic in AGN. The pressure-confining scenario requires
fragmentation of the confined phases into a large number of clouds.

\end{small}
% We
%have modelled the BeppoSAX data constraining the number of absorption
%components and the outflow velocity of these components with the
%results of the {\em Chandra} model.

\end{abstract}

%\keywords{galaxies: absorption lines -- galaxies: Seyferts --
%galaxies: active -- galaxies: X-ray}

\section{INTRODUCTION \label{intro}}

Highly ionized (or `warm') absorbers (Halpern 1982) have been
observed in about half
of all UV and X-ray spectra of AGNs (Reynolds
1997, George et al. 1998, Crenshaw et al. 1999). The absorption lines of these
systems are blueshifted with respect to the optical emission lines,
which implies outflow of material.  The mass loss rates
inferred from these outflows can be a substantial fraction of (Mathur,
Elvis \& Wilkes 1995) or even larger than (Behar et al. 2003, Ogle et
al. 2004) the
accretion rates needed to power the AGN continuum, indicating that
these `quasar winds' play an important role in the mass and energy
budgets of AGNs.

The UV and X-ray absorbers must be closely related, as exactly the
same AGNs show both \citep{ma95,ma98,cren99,mo01}.  Furthermore,
their kinematic properties also appear to be related, as both
absorbers show similar outflow velocities
($\sim$1000~km~s$^{-1}$). Given the intrinsic  complexity of UV
and X-ray spectra, along with the large difference in
spectral resolution in the two bands (with an order of magnitude
lower spectral resolution in the X-ray region), it has been
extremely difficult to connect these absorbers in detail, leading to
highly complicated models to describe them. Nevertheless, recent
studies using more comprehensive models have shown that the
observational data are certainly pointing to a common
nature for the UV and X-ray outflows (see \S \ref{uv}).

%which leave ionized absorbers as completely ineffective
%to study the circumnuclear environment of AGNs.

New studies based on high quality data have shown
that the absorbers can be described in a surprisingly simple way.
For example,
Krongold et al. (2003) modeled the 900 ksec {\em Chandra} HETG exposure
on NGC 3783 (Kaspi et al. 2002) and found that most of the absorption
features
present in the spectrum could be reproduced by a model with only
two absorption components, and that these components appeared to be
in pressure balance. This result was confirmed by Netzer et al. (2003), who
found a third highly ionized component, also in pressure
equilibrium with the other two. Is this single object result only a
coincidence?
Or, on the contrary, does it suggest that promising advances in
our understanding of the physical conditions of AGN winds can be
achieved, carrying important repercussions in our understanding of
the structure of quasars (e.g. Elvis 2000)?

Here, we present {\em Chandra} HETG and BeppoSAX  observations of the
Seyfert 1 galaxy NGC 985 and
show that the X-ray
spectra of this source can also be well described by a simple model
with two absorption
components. Furthermore, as for the ionized absorber toward NGC 3783,
these components are consistent with pressure
equilibrium, with one phase having sufficiently low ionization to
produce substantial absorption in the UV region. We further discuss
the implications of these results.

\subsection{NGC 985}

NGC 985 (Mrk 1048) is a Seyfert 1 galaxy located at redshift 0.04274$\pm
0.00005$ (12814$\pm 15$ km s$^{-1}$; Arribas et al. 1999, based on
stellar absorption features). NGC 985 is a peculiar galaxy with a
prominent ring shaped zone at several kiloparsecs from the nucleus
(de Vaucouleurs \& de Vaucouleurs 1975), suggesting that the galaxy
is undergoing a
merging process. Further evidence for a merger has been found
through optical and near-IR observations, which have revealed the
presence of a double nucleus (Perez Garcia \& Rodriguez Espinosa
1996, and references therein). The nuclei are separated by 3'' (2.5
kpc of projected linear distance; Appleton et al. 2002), which
indicates that the intruder
galaxy responsible for the formation of the ring is sinking
in the nuclear potential of the primary galaxy (Perez-Garcia \&
Rodriguez-Espinosa 1996). As in other galaxies
undergoing a merger, the IR luminosity of NGC 985 is extreme,
$L_{IR}=1.8\times 10^{11}~L_{\odot}$, which puts this object in the
LIRG group.

An ionized absorber has been observed in the X-ray spectra of
NGC 985. This source has been observed in the X-rays with different
satellites and is a strong X-ray source. The first detection was
with the Einstein Observatory, measuring an 0.2-4.0 keV
flux of 1.76$\times 10^{-11}$ erg s$^{-1}$ cm$^{-2}$ (a luminosity of
1.4$\times 10^{44}$ erg s$^{-1}$; Kruper, Urry, \&
Canizares 1990). Subsequently, spectral analysis of ROSAT-PSPC data
revealed complex features (Brandt et al. 1994). A simple power law
could not fit the PSPC spectrum of NGC 985, and  the presence of a
warm absorber, as well as a continuum soft excess, was suggested. In
1996, NGC 985 was observed by ASCA with a total exposure of 40
ksec. The analysis of these data confirmed the presence of an ionized
absorber in the spectra of this source (Nicastro et al. 1998,
1999). Nicastro et al. found that photoelectric bound-free
edge absorption could not account for all the features observed, and 
suggested that  resonant bound-bound line
absorption played an important role.

This possibility was supported by HST-STIS and FUSE ultraviolet
observations of the nucleus in NGC 985, which have revealed the
presence of six narrow absorption
components detected in \lia,  C {\sc iv}, N {\sc v} and O {\sc vi} (Arav
2002). Five of these  systems are observed in outflow, with
velocities (with respect to the centroid of the narrow emission
lines) of 780, 670, 519, 404,
and 243, \kms (hereafter components 1 to 5). The sixth component is
observed inflowing at $-$140 \kms. As demonstrated by Arav, small to moderate
saturation is present in the \lia \ trough of these
components.  Once the effects of saturation are accounted for, at
least component 4 is consistent with gas that also produces X-ray
absorption, i.e. a highly ionized or `warm' absorber, with a column
density of 10$^{21.4}$ cm$^{-2}$.

In this paper we present 2 moderate exposure observations of NGC 985
($\sim$ 100 ksec each) with {\em Chandra} and BeppoSAX. In \S \ref{dat} we
describe the reduction
and spectral analysis of the {\em Chandra}-HETGS data. The analysis was
carried out in Sherpa \footnotemark\ (Freeman, Doe, \& Siemiginowska
2001), with the inclusion of the code  PHASE (Krongold et al. 2003)
to model in a self-consistent
way the ionized absorber present in the spectra.  In \S \ref{sax_dat}
we present the BeppoSAX data, and in \S \ref{vari} we analyze the
variability effects of the source on the absorber. In \S\S \ref{disc}
and \ref{col} we discuss the different scenarios consistent with the
absorber and their implications.

\footnotetext{http://cxc.harvard.edu/sherpa}

%The errorbars throughout this paper
%correspond to the 2$\sigma$ level, except where otherwise stated.

\section{THE {\em Chandra}-HETG SPECTRUM \label{dat}}

On 2002 June 24, the  High Energy Transmission Grating Spectrometer
(HETGS, Canizares et al. 2000) on board the {\em Chandra} X-ray
Observatory (Weisskopf et al. 2000)
observed NGC 985 (see Table \ref{log}). The
observation was carried out with the Advanced CCD Imaging Spectrometer
(ACIS; Garmire et al. 2003) in the focal plane,
and had a duration of 80 ksec.  We
processed the event files of this observation with the {\em Chandra}
Interactive Analysis of Observations (CIAO\footnotemark
\footnotetext{http://asc.harvard.edu/ciao}; Fruscione 2002) software
(Version 2.3), following the on-line data analysis ``threads'' provided by the
{\em Chandra} X-ray Center (CXC\footnotemark
\footnotetext{http://asc.harvard.edu/ciao/documents\_threads.html}), and
extracted source and background 1st order spectra
(positive and negative orders) and responses for the High-Energy
Grating (HEG) and Medium-Energy Grating (MEG) configurations.  We
accounted for the ACIS efficiency degradation (Marshall et
al. 2003) using the script {\it
contamabs}\footnotemark \ (version 1.1\footnotetext{http://space.mit.edu/CXC/analysis/ACIS\_ Contam/ACIS\_
Contam.html}). The net exposure of the observation is 77.7 ksec and
the total number of counts in the dispersed spectrum is 13,979.

\begin{figure}
\plotone{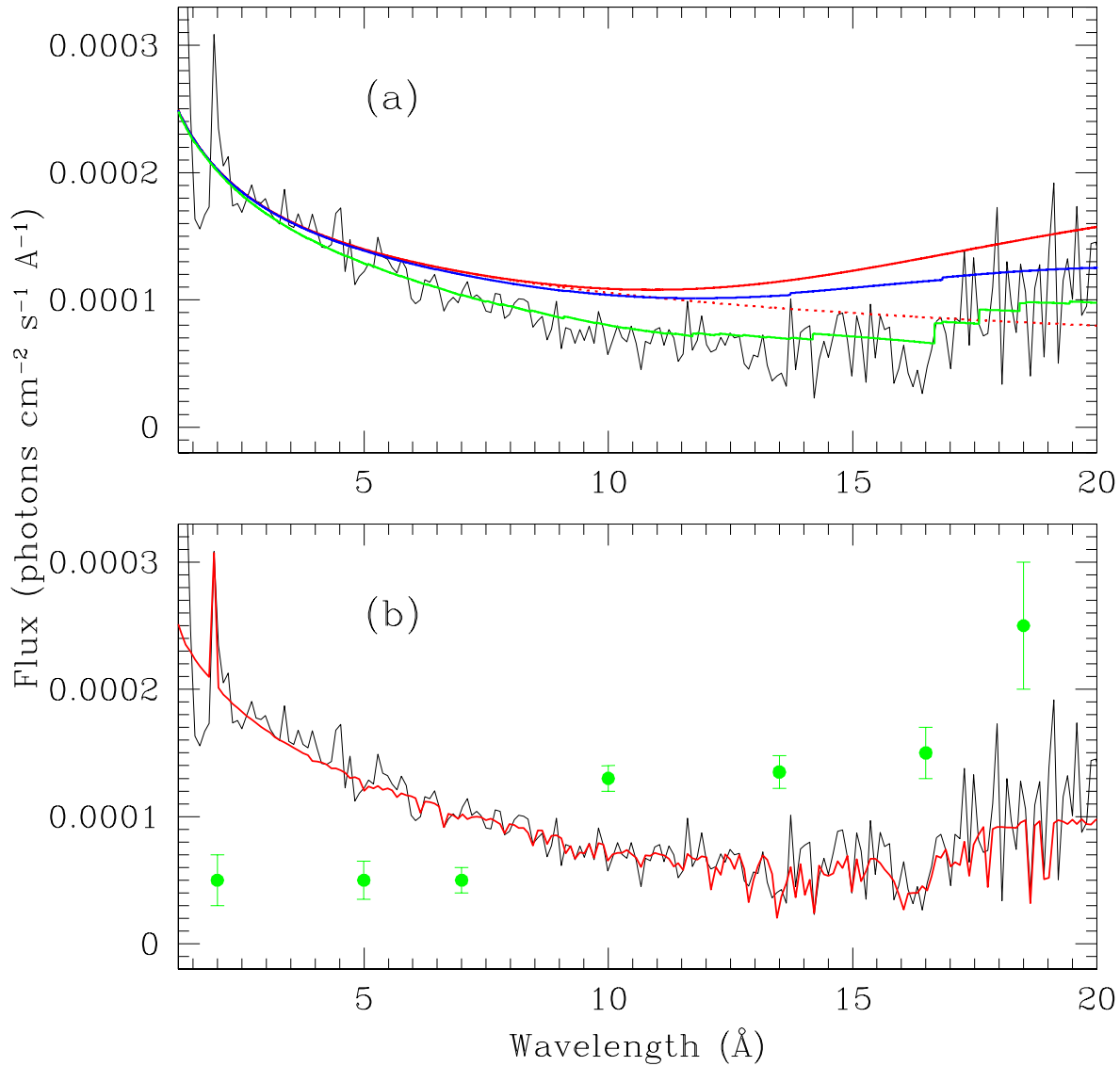} \caption[f1.eps]{\footnotesize{Fluxed, combined (MEG+HEG)
first-order spectrum of NGC 985. (a) Continuum fitted to the
spectrum. The three solid lines stand for (from upper to lower):
intrinsic continuum of the source, continuum attenuated by
Galactic absorption, and continuum further attenuated by the
bound-free photoelectric absorption edges. The dotted line represents
the predicted power law without the contribution from the blackbody
component.
(b) Two phase absorption model plotted for comparison. Typical errors
(per bin) are marked with green solid ranges.{\it [See the electronic
edition of the Journal for a color version of this figure]}} \label{flux}}
\end{figure}

We then combined the -1st and +1st order spectra to improve the signal
to noise (S/N) ratio of the data. Similarly, after checking for
wavelength scale consistency, the HEG and MEG spectra and
their responses were
coadded in the overlapping region, to further increase the S/N ratio.
The S/N ratio per MEG resolution element
($\approx$0.02 \AA) varies from $\approx 7$ to $\approx 2$ in the
2-20 \AA \ range, a factor $\approx 6$ lower than the highest S/N ratio
HETG spectrum of a Sy 1 galaxy (NGC 3783; Kaspi et al. 2002).
All the spectra presented throughout the paper
(except where otherwise stated)
correspond to a binning factor of 20 (i.e we binned the spectrum in
bins of size 0.1 \AA, about 5 times the MEG resolution
element), giving a S/N ratio ranging from 14 to 3 per bin (in the 2-20 \AA \
range). The spectrum is shown in
Figure  \ref{flux}a.
%Figure \ref{flux}a shows the unfolded spectrum, and the final
%empirical spectrum is shown in Figure \ref{spectra}.

\subsection{Spectral Analysis \label{fit}}

%We performed spectral fitting of the NGC 985 data using the code PHASE
%(Krongold et al. 2003) integrated into the Sherpa\footnotemark \ (Freeman et
%al. 2001) package  in CIAO (Fruscione 2002). \footnotetext{http://cxc.harvard.edu%/sherpa}

In the calculations reported here, we attenuated the continuum by an
equivalent hydrogen column density of $3.0\times10^{20}$ cm$^{-2}$
to account for the Galactic absorption by cold gas
in the direction to the source (Starck et al. 1992). We have explored
only photoionization equilibrium models, and we have assumed solar
elemental abundances (Grevesse \& Noels 1993).

\subsubsection{Fitting the Continuum \label{chandra_cont}}

\begin{figure}
\plotone{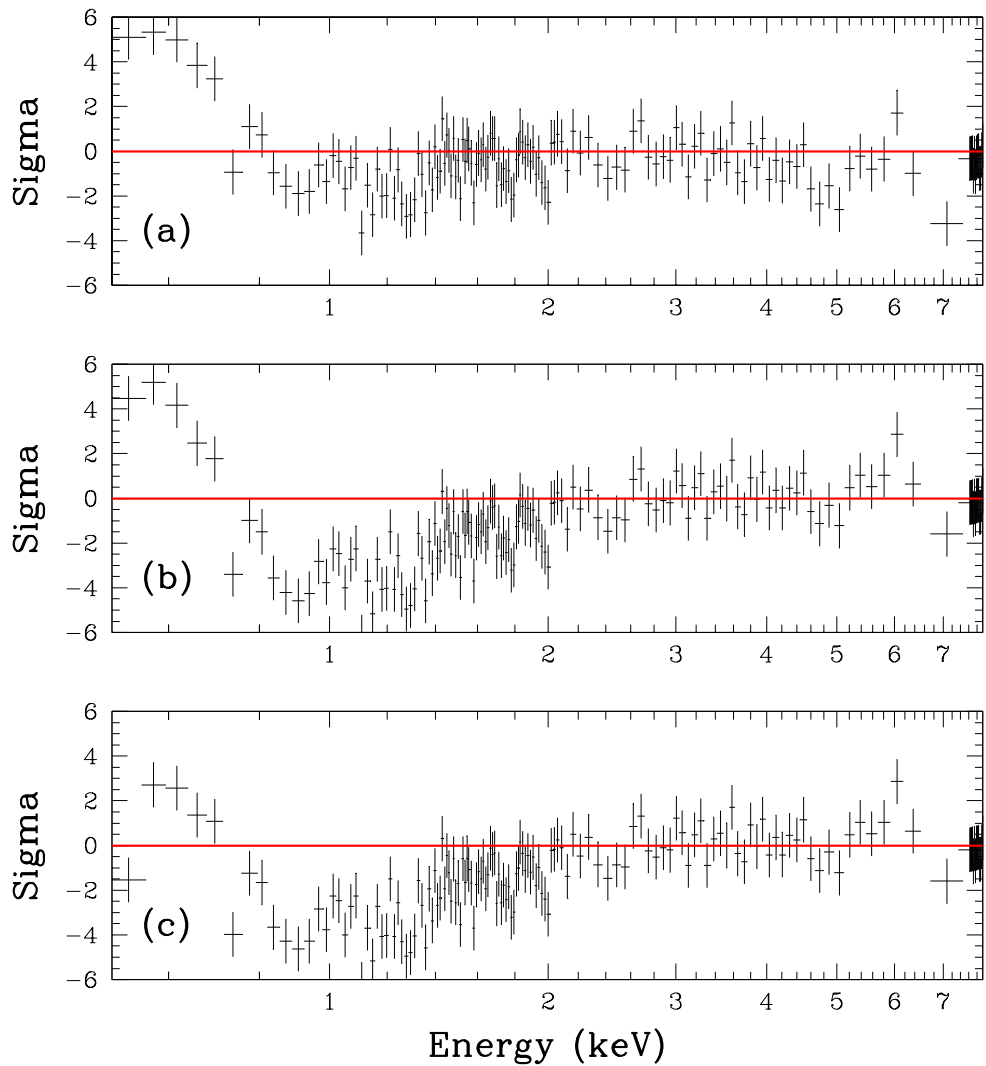} \caption[f2.eps]{Fit residuals
for continuum determinations over {\em Chandra} HEG+MEG data of
NGC 985. All models are attenuated by cold gas with
$\log$N$_H=20.5$ [cm$^{-2}$] due to  Galactic absorption.
(a) Model including
a continuum power law fitted over the entire spectral range
(0.6-8keV). (b) Power law fitted only in the 3-8 keV range. (c)
Model in Panel (b) plus a soft excess parameterized with a
blackbody component. The data are binned to 
$>$ 100 photons in each
channel to make evident the overall features.
{\it [See the electronic edition of the
Journal for a color version of this figure]} \label{res_chandra}}
\end{figure}

To model the intrinsic continuum of the source, we first used a
simple power law. This model could not fit the data over the entire
energy range (0.6-8 keV, 1.5-20 \AA), as can be seen from the residuals
(Fig. \ref{res_chandra}a). The model shows both positive and negative
deviations (with significance $> 2 \sigma$) typical of the ionized
absorber and soft excess commonly observed in Seyfert 1 galaxies
(e.g. Reynolds 1997).
% A lot  of these residuals are significant  and
%cannot be
%attributed to calibration uncertainties. Rather, these features
%make
%evident the presence of both an ionized absorber and a soft excess.

We repeated the power law fit using only data
between 3 and 8 keV (we also excluded the Fe K$\alpha$ line region
around 6.4 keV). At these energies, absorption by moderately large
columns of ionized gas (i.e. $<10^{23}$ cm$^{-2}$) or low columns of
neutral gas (i.e. $<10^{21}$ cm$^{-2}$, Galactic gas) do not
modify significantly the shape of the intrinsic continuum (e.g.
Krongold et al. 2003). The results of this model are summarized in
Table \ref{tabcont}, and the residuals (extrapolated to the whole
spectral range) are presented in Figure \ref{res_chandra}b. The
presence of strong positive and negative deviations below 3 keV in
this model shows that both an ionized absorber and a soft
excess are likely to be present. To account for the presence of
the excess, we further
included a blackbody component in our continuum determinations,
and the positive residuals decreased significantly (see Fig.
\ref{res_chandra}c). However, the model still shows strong
negative residuals at energies characteristic of an ionized absorber.

Finally, we fitted the continuum and the absorption together (see
\S \ref{phase} for the details of the ionized absorber). In this
model we left all the continuum components (power law
and blackbody) free to vary, and we assumed that the absorbing gas
also covers the soft excess. The results obtained for the power
law with this
approach are in excellent agreement with the results obtained when
considering only the 3-8 keV range.
% We also find that the soft
%excess can, indeed, be parameterized appropriately by a blackbody
%component.
Table \ref{tabcont} presents the values for the best
continuum fit. Figure \ref{flux}a shows the fluxed data over the
entire spectral range (1.5-20 \AA, 0.6-8 keV), together with (a)
the intrinsic continuum of the source, (b) the continuum
attenuated by Galactic absorption, and (c) the observed continuum
(i.e. the continuum further attenuated by the bound-free
transitions from the ionized absorber, see details in Fig.
caption).

\subsubsection{Modeling the Ionized Absorber \label{phase}}

%We used PHASE (Krongold et al. 2003) to model the absorption features
%present in the spectra of NGC 985,

\begin{figure}
\plotone{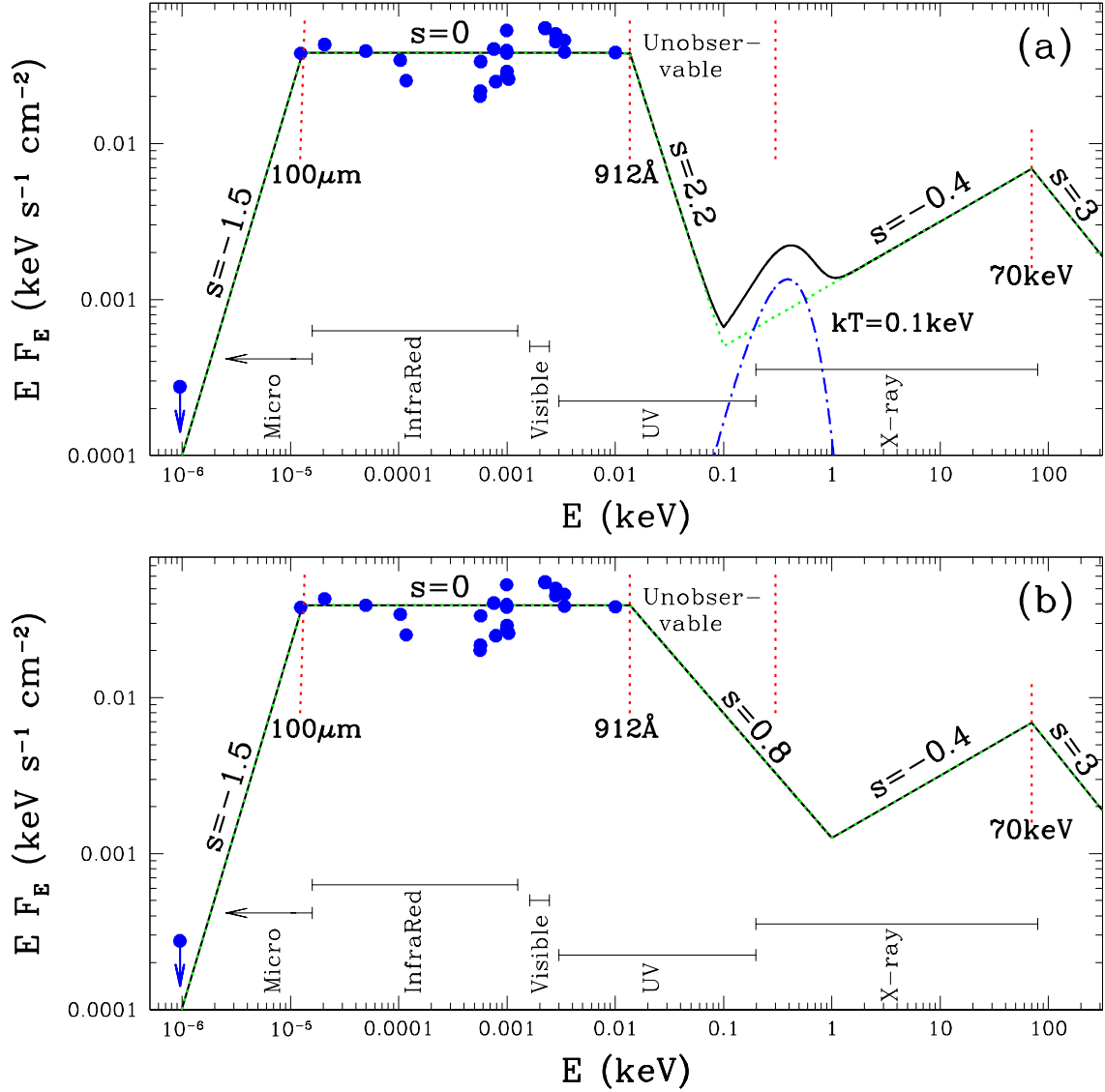} \caption[f3.eps]{Spectral Energy
Distribution used to model the ionized absorber in the {\em
Chandra}-HETG data. Filled circles (blue) mark the observed SED of
NGC 985 (obtained from NED). The plot shows the slope adopted in each
energy range. This slope relates to the photon index as
$\Gamma$=s+2. (a) Model
with low energy far UV cutoff at 0.1 keV and including the
contribution of a blackbody component. (b) Model with high
energy far UV cutoff at 1 keV. {\it [See the electronic edition of the
Journal for a color version of this figure]} \label{fsed} }
\end{figure}

We ran PHASE with three free parameters for each component
necessary to fit the ionized absorber: (1) the ionization
parameter (defined over the entire Lyman continuum, as
$U=Q(H)/4\pi r^2 n_H c$ with $Q(H)$ being the rate of H ionizing
photons, $r$ the distance to the source, $n_H$ the H
density, and $c$ the speed of light.), (2) the equivalent hydrogen
column density N$_H$, and (3) the outflow velocity. Other
parameters of the code that we kept fixed are the internal
micro-turbulent velocity of the absorbers and the intrinsic
Spectral Energy Distribution (SED) of the source (Fig. \ref{fsed}a).
The turbulent
velocity of the gas is very difficult to constrain, since (1) single
absorption lines in the spectrum are unresolved by
the HETGS (which has a FWHM resolution of $\approx$ 400 \kms at 17
\AA) and (2) most of the observed features are blends of several
transitions. We set this parameter to 350 \kms, the value most
suitable to the data after several tests comparing the measured
equivalent widths (EWs) of a few unblended lines (Si {\sc xiv}, Si
{\sc xiii}, Mg {\sc xii}, Mg {\sc xi}, Fe {\sc xvii}) with the EWs
predicted by our model.

%Mathur et al. (1994) showed the importance of including the particular
%SED of an AGN, rather than a typical SED when modelling spectroscopic
%data. Otherwise, incompatible physical conditions may arise in the
%description of a photoionized cloud.
To approximate the SED, we proceeded as follows: in the X-ray
region we used the power law $\Gamma$, as well as the blackbody
contribution, as inferred from the fits to the continuum presented
in \S \ref{chandra_cont}. To obtain the exact parameters for the
blackbody component several iterations were performed. The BeppoSAX
data of NGC 985 show the presence of a high energy cutoff in
the intrinsic continuum (see \S \ref{sax_cont}), which we modeled
with a break in the X-rays at around 70 keV, and a powerlaw with
$\Gamma=5$ above this energy.
Based on HST-STIS
observations of NGC 985, Bowen, Pettini \& Blades (2002) found a
flux of 2.54$\times 10^{-26}$ erg s$^{-1}$ cm$^{-2}$
hz$^{-1}$ at 1230 \AA. In order to determine the flux at the Lyman
limit, we extrapolated to 912 \AA \ assuming a power law with
photon index $\Gamma =$2. The flux at 912 \AA \ turned out to be
1.83 $\times 10^{-26}$ erg s$^{-1}$ cm$^{-2}$ hz$^{-1}$.
Extrapolation of the X-ray continuum down to the Lyman limit
gives a flux two orders of magnitude smaller than measured by HST, which
indicates the presence of a soft X-ray-UV energy break, with
a steepening of the spectrum between the two bands (Fig. \ref{fsed}a).
We assumed
that this break occurs at 0.1 keV.  The actual energy of this
break is unknown (due to the unobservable region in the UV-X-ray
range associated with Galactic absorption).
We restricted the SED at low energies
using (not simultaneous) data from NED\footnotemark\ (see Fig.\ref{fsed}a).
For simplicity we extended
the extrapolation between 1230 \AA\ and 912 \AA\ up to 100 $\mu$m (i.e.
we considered a power law with $\Gamma =$2 from 912 \AA\ to 100
$\mu$m). Below 100 $\mu$m we introduced
a low energy cutoff with $\Gamma=-3.5$.
The final SED used to model the {\em Chandra}
data is presented in Figure \ref{fsed}a. The differences in the
infrared region between NED data and our assumed SED do not have any
effect on our analysis (the only effect such differences produce is an
increase of the Compton temperature by a factor $<$1.1, when the NED
data are considered).

%and reflects the properties
%of the observed SED of NGC 985 (data obtained from NED
%for low energy ranges).

\footnotetext{NASA/IPAC Extragalactic
Database, http://nedwww.ipac.caltech.edu}

\subsubsection{The Ionized Absorber \label{abs_chandra}}

\begin{figure}
\plotone{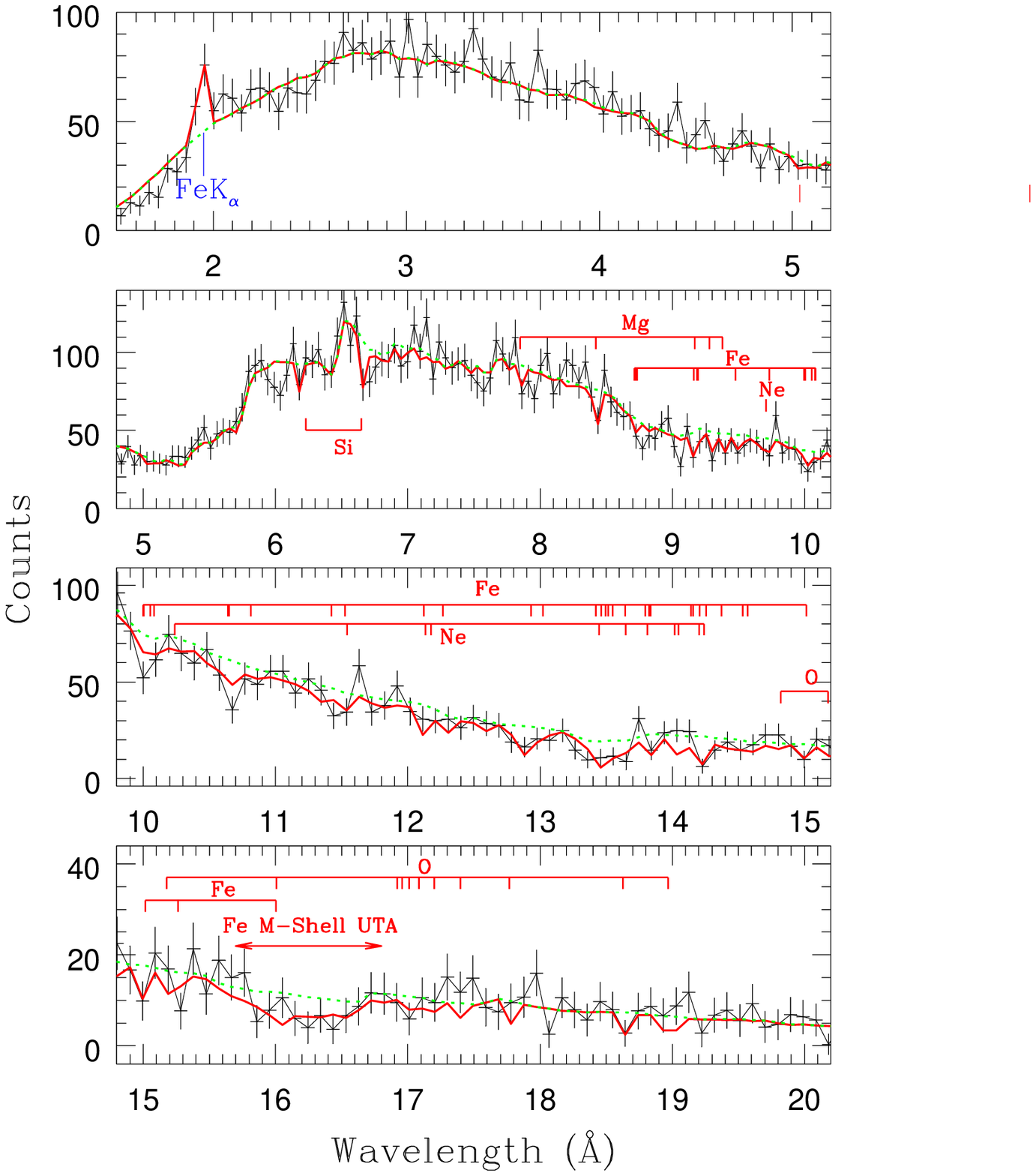} \caption[f4.eps]{Two phase absorber
model plotted against the first-order combined MEG+HEG spectrum of
NGC 985. Absorption lines predicted are marked 
along bars at top for each element (red).
The continuum level (including edge continuum absorption) is
overplotted for comparison (dotted green line). The spectrum is
presented in the rest frame system of the absorbing gas. Below 10 \AA\
the data are presented in bins of size 0.05 \AA, above  10 \AA\ in bins
of size 0.1 \AA.
{\it [See the electronic edition of the
Journal for a color version of this figure]} \label{spectra}}
\end{figure}

\begin{figure}
\plotone{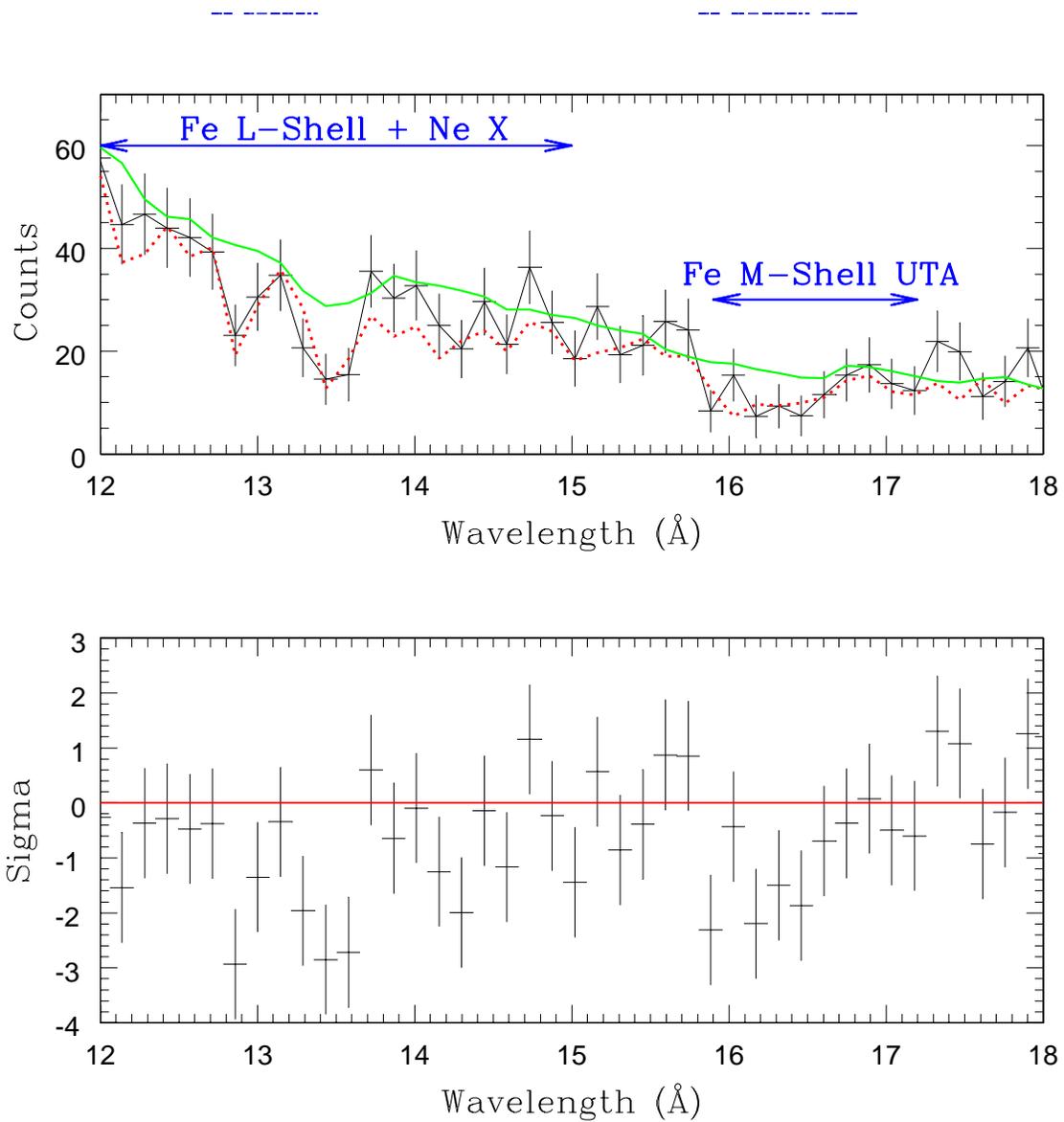} \caption[f5.eps]{ \footnotesize Upper panel:  MEG+HEG
 spectrum of NGC 985 grouped in bins of size 0.15 \AA \
(binning factor of 30). The
solid line (green) marks the
continuum level, including photoelectric
bound-free absorption from our model. The final model including
resonant absorption is marked with the dotted line (red). The
feature in the continuum (solid green line) between 13.3 and 13.8 \AA\
is produced by a chip gap in the detector. Lower panel:
Residuals (data minus continuum green solid line) in units of standard
deviations showing the presence of deep 
Fe L-shell resonant absorption (significance $\approx 3 \sigma$ per bin) and 
Fe M-shell inner shell resonant absorption (significance $\approx
2 \sigma$ per bin).{\it [See the electronic edition of the
Journal for a color version of this figure]}
\label{delchi}}
\end{figure}

As is the case 
for other ionized absorbers in AGNs (Kaspi et al. 2001; Blustin et. al
2002; Krongold et al. 2003; Netzer et al. 2003 for NGC
3783; Kaastra et al. 2002 for NGC 5548; Steenbrugge et
al. 2003 for NGC 4593; Blustin et al. 2003 for NGC 7469) the ionized
absorber in the
spectrum of NGC 985 could not be well fitted by a single absorbing
component ($\chi^2/$d.o.f.$=201.6/177$). Therefore, we attempted to fit
the spectrum with the
next simplest model, by including
a second component. In this case, we obtained an acceptable fit to the
data 
($\chi^2/$d.o.f.$=179.03/174$; an F-test indicates that the second
component is required at a significance level larger than
99.9\%). Table  \ref{2ab} lists our results, and 
the best fitting model is presented in Figure \ref{spectra} (and
also in Fig. \ref{flux}b, where we show the model for the full
spectral range plotted over the fluxed spectrum). Most of the narrow features
observed in  Figure \ref{spectra} have a significance $<2\sigma$, but
still the absorber is clearly detected and constrained through the
presence of relatively broad features due to both photoelectric
bound-free absorption and blends of resonant bound-bound transitions
(i.e. Fe L-shell and Fe M-shell absorption).
To highlight
the presence of these features (which are significant
at a level $>2.5\sigma$), in Figure
\ref{delchi} we present our model and the data in the 12-18 \AA \ range,
with a binning size of 0.15 \AA \ (binning factor of 30).

\begin{figure}
\plotone{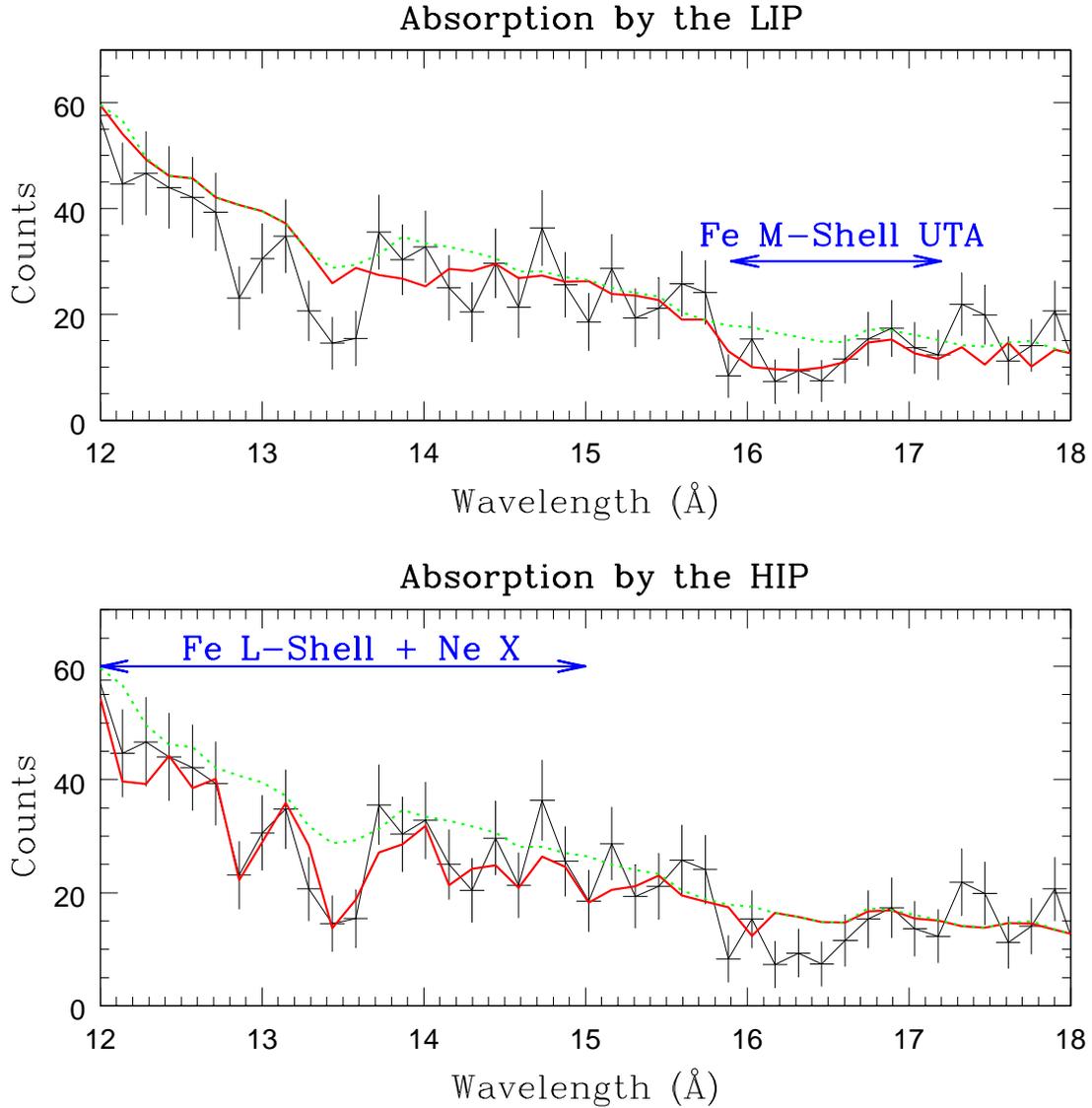} \caption[f6.eps]{Upper Panel:
Resonant bound-bound
absorption produced
by the low ionization phase (LIP) of our model, plotted against the
MEG+HEG spectrum of
NGC 985 grouped in bins of size 0.15 \AA \ (binning factor of 30).
The dotted line (green) marks the
continuum level, including bound-free photoelectric absorption from
our model.
The main feature of this component is the Fe UTA. Lower panel: As
upper panel, but showing the contribution from the high ionization
phase (HIP). Below 15 \AA \ the spectrum is dominated by 
Fe L-shell
absorption.{\it [See the electronic edition of the
Journal for a color version of this figure]}
\label{2comp}}
\end{figure}

\begin{figure}
\plotone{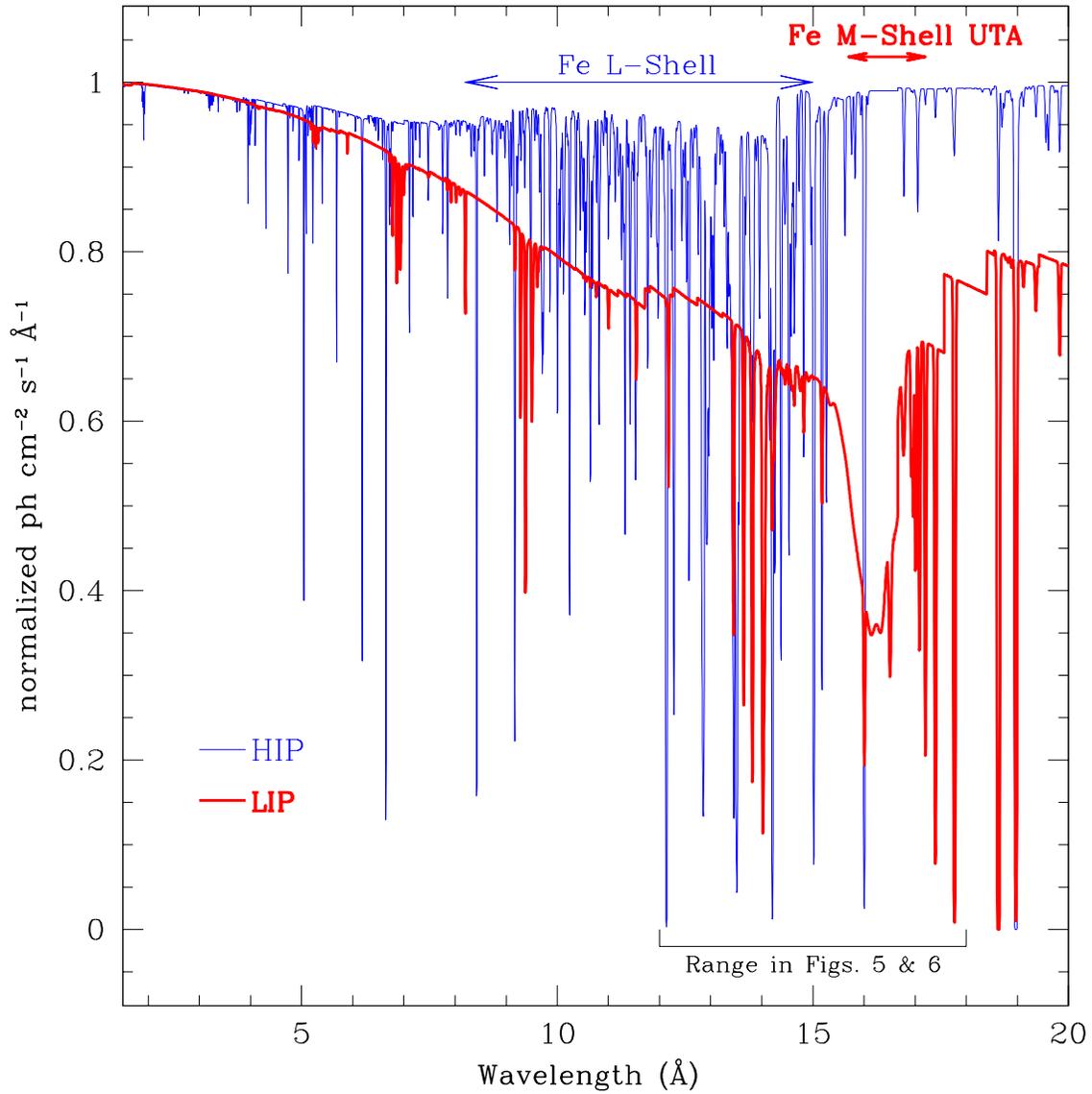} \caption[f7.eps]{Theoretical transmission spectrum
of the two absorption components at 0.001 \AA \ resolution. The low
ionization component produces
most of the bound-free photoelectric absorption and the UTA near
16 \AA. The high ionization component produces the 
Fe L-shell
absorption below 15 \AA. {\it [See the electronic edition of the
Journal for a color version of this figure]} \label{model}}
\end{figure}

Two different ionization absorbing components are clearly
required by the data. The presence
of significant features produced by both Fe L-shell
transitions
(Fe {\sc xvii-xxii}) and Fe inner M-shell
transitions (Fe {\sc vii-xiii}, the unresolved transition array or UTA)
is incompatible with a single ionization degree for the
absorbing gas. This is shown in Figure \ref{2comp},
where the
contribution from the two components to the absorption is presented
separately. The high ionization phase (hereafter
HIP), with $\log$U=1.34, gives rise to the absorption by \oviii, Ne
{\sc ix-x}, Mg {\sc xi-xii}, Si {\sc xiii-xiv}, and Fe L-shell.
The low ionization
phase (hereafter LIP), with $\log$U=-0.12, is responsible for the
absorption by charge states O
{\sc vii}, Ne {\sc v-ix}, Mg {\sc ix}, and the Fe M-shell UTA.
The components are well
separated in ionization degree, the HIP has a value $\approx29$
times larger in U than the LIP.  Accordingly, the HIP has a higher
temperature ($\log$T=5.8 [K])
than the LIP ($\log$T=4.5 [K]) by a factor of 21. These temperatures
were calculated
assuming photoionization equilibrium in each component. Figure
\ref{model} shows the theoretical spectral features predicted for the
absorption components. Table \ref{ionic} presents the predicted column
densities for the dominant charge states in the LIP and the HIP.

\begin{figure}
\plotone{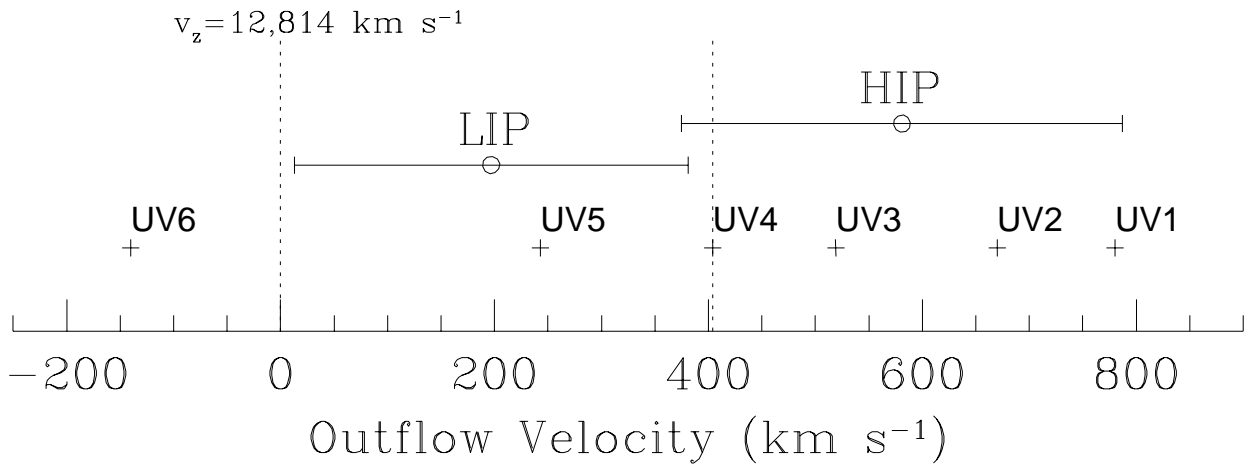} \caption[f8.eps]{Outflow velocities for the HIP
and the LIP, and for the 6 absorption components observed in the
UV. The HIP is consistent within 2$\sigma$ with UV components 1
to 5, and the LIP with components 3 to 6. Both the HIP and the LIP
are consistent with component 4, which was expected to absorb in
the X-ray region (Arav 2002). \label{vels}}
\end{figure}

As has been observed in
other ionized absorbers (see above), the HIP has a larger equivalent H
column density than the LIP, in the case of NGC 985 by a factor
of 3. The outflow velocity of the components is 581$\pm
206$ \kms \ for the HIP and 197$\pm 184$ \kms \ for the LIP.
These velocities, though ill-determined, are consistent
with each other (within $\sim 1 \sigma$) and with those observed in the
HST and FUSE systems (see Fig. \ref{vels} for more details).
A fit fixing the
velocity of both HIP and LIP at the velocity of the strongest UV
component (4, outflowing at 404 
\kms; see \S \ref{intro} and Fig. \ref{vels})
yields an acceptable fit to the data, with parameters fully consistent
with those obtained before.

In the former analysis we have assumed that the soft excess detected
in the data is covered by the absorber. Soft X-ray excesses have been
studied in low resolution spectra for more than a decade 
(e.g., Elvis et al. 1991, and references therein) and they have
been found to vary 
on timescales of tens of days. This effect has been confirmed in high
resolution spectra (Netzer et al. 2003), and suggests that the
bulk of the soft X-ray emission comes from a compact region (less than
$10^{16}$ cm 
across; Elvis et al. 1991). It has been suggested that
the emission might
have a thermal nature and might originate in the inner edges of
accretion disks (e.g. Czerny 
\& Elvis 1987). Therefore,
it is reasonable to assume that the absorbing material lies further
away from the central engine than the soft X-ray
emission. Nevertheless, we note that the data does not allow to
distinguish whether the soft component is covered by the absorber or
not. A fit assuming an uncovered excess is statistically
indistinguishable from a fit which assumes full covering. If the
excess is not absorbed, then the column density of the LIP is 1.6 times 
larger and the normalization of the blackbody 2.3 times
smaller. The rest of the parameters in this model are fully consistent
with the values obtained before.  We stress that whether the soft
excess is covered or not covered by the absorber, this has 
a negligible effect 
on the conclusions presented in this paper.

We find no significant evidence of an ionized emitter in the
spectrum. Resonant and forbidden \ovii \ lines have a
statistical significance $\lesssim$ 1 sigma. We estimate upper limits
for the equivalent width of 218 m\AA \ for the resonant line and 185
m\AA \ for the forbidden line.
Although an emitter might
be present, better data are required for its detection.

\subsection{Constraining the Ionized Absorber \label{absfits}}

Despite the limited S/N ratio of the {\em Chandra} spectra, the physical
properties of the ionized
absorber in NGC 985 can be constrained. As
demonstrated by Behar, Sako, \& Kahn (2001) and Krongold et
al. (2003), the UTA is a robust indicator of the ionization level of
the gas, as the general shape and position of this feature strongly
depend on 
U for a given ionization component. The 12-15 \AA \ Fe L-shell absorption
complex also helps to
limit the physical state of the gas.

\begin{figure}
\plotone{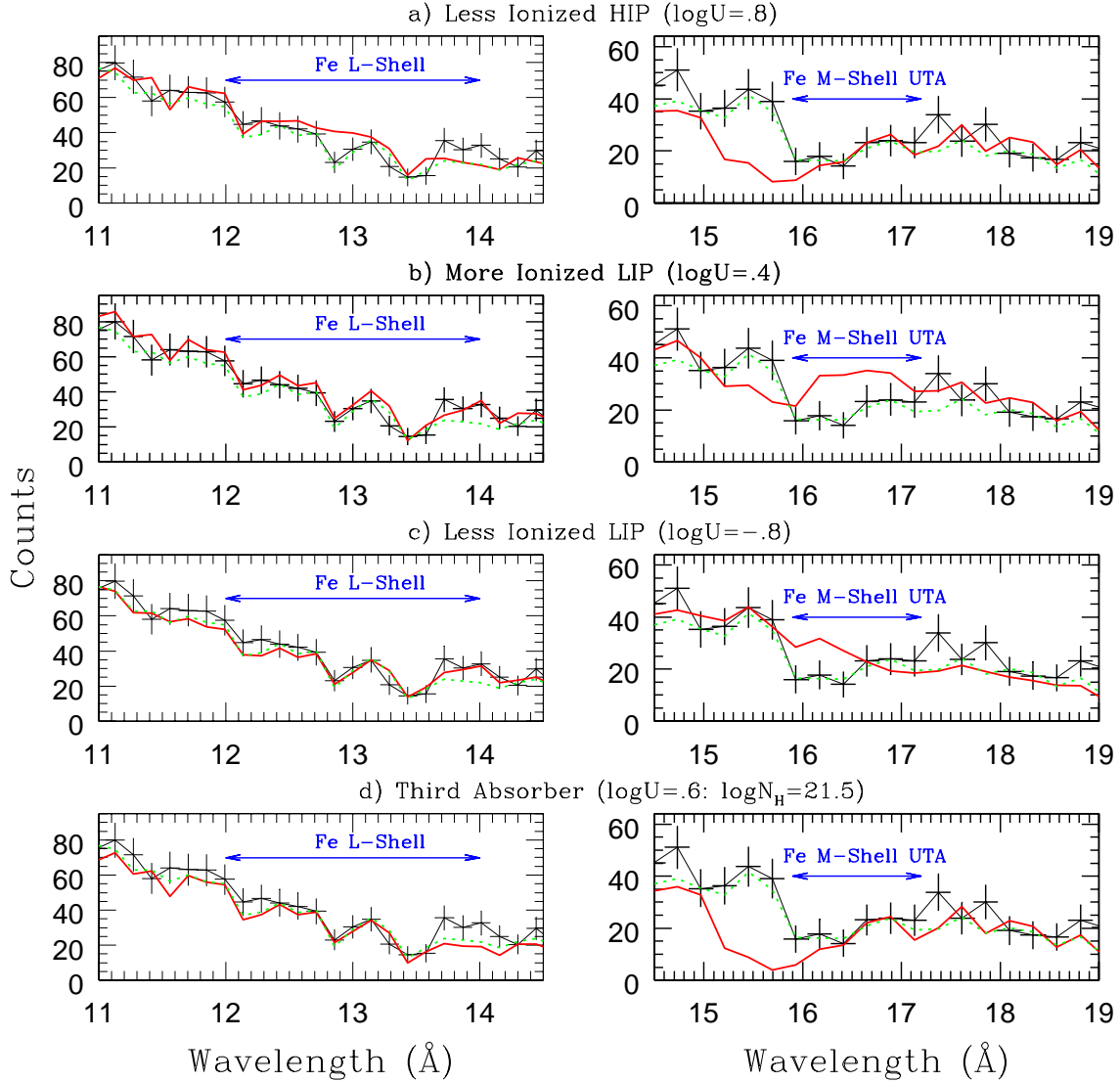} \caption[f9.eps]{ \footnotesize Different
values for the ionization parameters of our two phase absorber
showing inconsistencies with the observed UTA. The left panels present
the data in the range 11-14.5 \AA \ in bins of size 0.15 \AA \
(binning factor of 30), the
right panels the data in the range 14.5-19 \AA \ in bins of size 0.25
\AA \ (binning factor of
50). The green dotted line shows the best fit model for comparison.
Panel (a): Less
Ionized HIP, $\log$U=0.8. Note that the absorption from Fe L-shell
is underestimated.  Panel (b): More ionized LIP, the ionization
parameter of the low ionization component has been set to
$\log$U=0.4. Panel (c): Less Ionized LIP, $\log$U=-0.8. Panel
(d): A third absorber has been included in our model with
$\log$U=0.6 and $\log$N {\sc h}=21.5 [cm$^{-2}$]. 
Including a new component 
overpredicts the absorption by the UTA, which further
suggests that
the absorbing gas tends to clump. {\it [See the electronic edition of the
Journal for a color version of this figure]} \label{fits} }
\end{figure}

To illustrate this point, we have constructed a series of models
with ionization parameters different from those deduced for our
best fitting model (see \S \ref{abs_chandra} and Table \ref{2ab}).
The comparison of these models with the data reveals clear
discrepancies (see Fig. \ref{fits}).
The first model we considered
is presented in Figure \ref{fits}a, and consists of a factor 3 less ionized
HIP ($\log$U=0.8). In this model, lower charge states of Fe would
be significantly present in the HIP (e.g.
Fe {\sc xv-xvi}), and these would produce abundant 
Fe M-shell UTA
absorption at wavelengths between 15 and 16 \AA \ (see Fig.
\ref{fits}a). This is clearly not observed in the data. At the
same time, the deep absorption feature observed in the data around
12.85 \AA \ (arising from Fe {\sc xix-xx}), is missing in the
model. Thus, the HIP cannot contain less ionized gas.

The ionization state of the LIP is also limited by the data,
particularly by the UTA. Figure \ref{fits}b shows the comparison
of the data with a second model that includes a factor 4 more ionized LIP
($\log$U=0.5). This model shows an excess of absorption at 15- 16
\AA \ produced by Fe {\sc xiv-xv}, and a lack of absorption at
longer wavelengths (due to a diminished fraction of Fe in charge
states {\sc vii-x}). A third model consisting of a factor 5 less ionized
LIP ($\log$U=-0.8) is presented in Figure \ref{fits}c. This model
is inconsistent with the data also. A low value of U shifts the
UTA to higher wavelengths (peak of absorption at 17 \AA) because
an important fraction of Fe is now present in charge states {\sc
iv-vi}. At the same time, the lack of Fe {\sc x-xii} produces an
underestimation of the absorption around 15.9-16.4 \AA. From this
analysis is evident that  even lower values of U for the LIP
($\log$U$<$-0.8) would produce even more shifted UTAs (produced by
even lower Fe charge states), and therefore, higher discrepancies
between data and model. Thus, the data impose important
restrictions on the ionization state of the absorbing gas.

A third absorber with an intermediate value of the ionization parameter
(between the HIP and the LIP) and high H column density can also be
ruled out. Figure \ref{fits}d illustrates the inconsistencies that
arise when a third absorber with $\log$U=0.6 and $\log$N$_H$=21.5
[cm$^{-2}$] is
included in our model. Such an absorber severely overpredicts the
absorption by the UTA from 15 to 16 \AA \ (because of a high
contribution to the absorption in the model by Fe
{\sc xiv-xvi}). However,  it is not possible to
rule out the presence of additional absorbing components with lower H
column density.

\begin{figure}
\plottwo{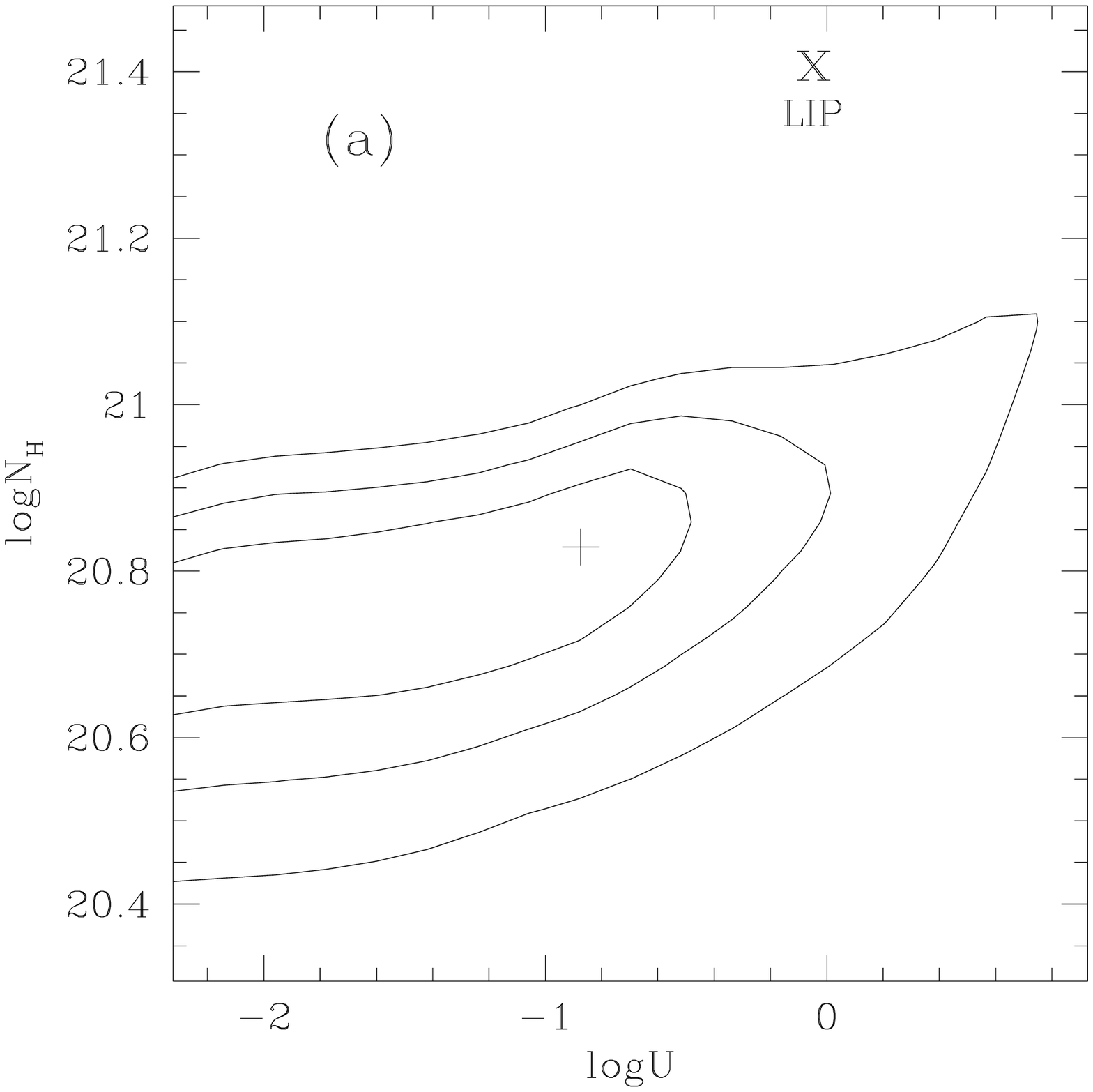}{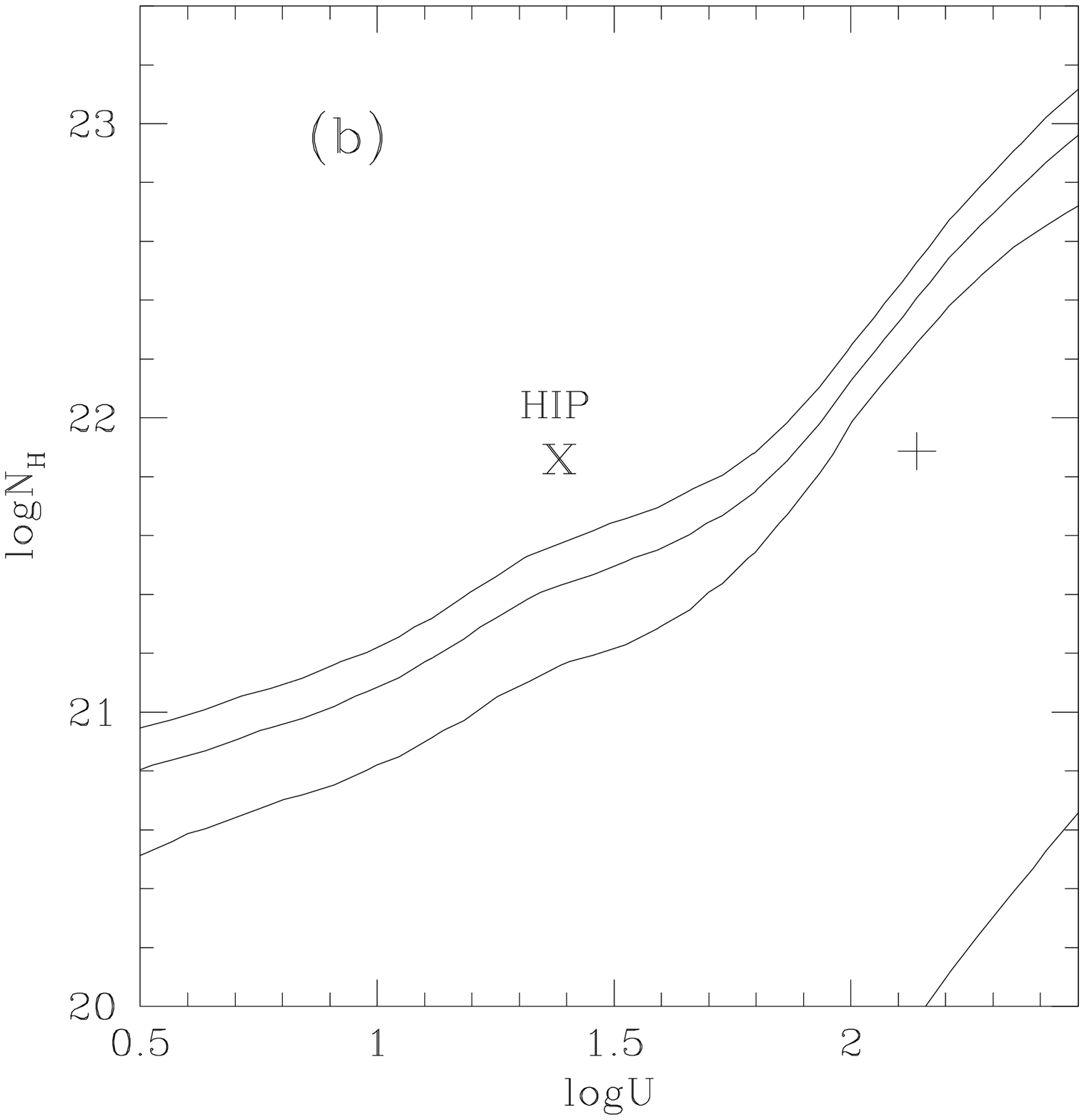}
\caption[f10.eps]{Confidence regions constraining the presence
of a third absorber in the {\em Chandra} spectrum of NGC 985. The
solid lines correspond to the 1$\sigma$,  2$\sigma$, and 3$\sigma$
levels.  Panel (a): A low ionization absorber with low column density
($\log$U=-1.0 and $\log$N{\sc h}=20.8 [cm$^{-2}$]).
 Panel (b): A high ionization component with high column ($\log$U=2.1
and $\log$N{\sc h}=21.8 [cm$^{-2}$]).
\label{cont}}
\end{figure}

To quantify the limits imposed by the data to additional absorbing
material we included
a third absorption component in our model and refit the data. In
Figure \ref{cont}a we
show the $\log$U-$\log$N$_H$ confidence region for this
hypothesized absorber. The plot shows that gas with U between the HIP and the LIP
and $\log$N$_H>21.1$ [cm$^{-2}$] (half the
column density of the LIP and 20\% that of the HIP) can be ruled out
(at the 3$\sigma$ level). Figure \ref{cont}a also shows that the
presence of material less ionized than the LIP and $\log$N$_H>21$
[cm$^{-2}$]
can also be ruled out (at 3$\sigma$ significance).

\begin{figure}
\epsscale{.5}
\plotone{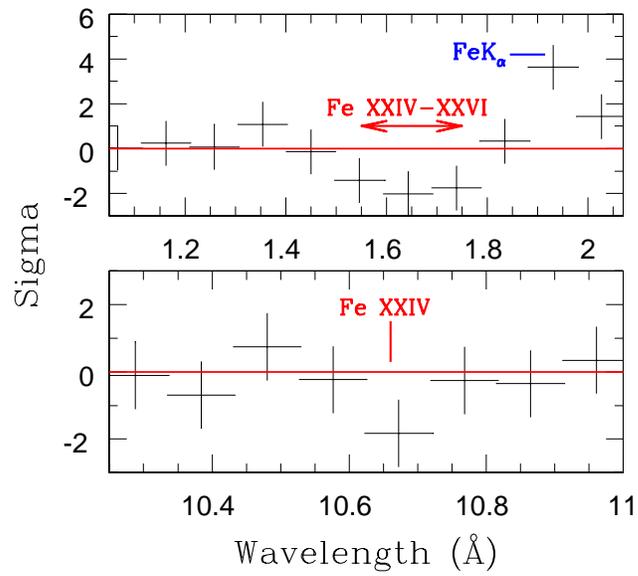} \caption[f11.eps]{Residuals to our best fit
model over {\em Chandra} data in the ranges 1.1-2 \AA \ and
10.2-11 \AA. The presence of absorption by material with
significant amounts of Fe {\sc xxiv-xxvi} is suggested in the
figure, but better S/N ratio is required to constrain the physical
properties of such component. {\it [See the electronic edition of
the Journal for a color version of this figure]} \label{res_high}}
\end{figure}

On the other hand, a third component with very high ionization
parameter and considerable
column density can be present in the spectra of NGC 985.
Figure \ref{cont}b presents the confidence region for a 3rd component
consisting of high column density and high ionization gas. As observed
in the figure such 
a component is not well constrained by the data, but
in order to have larger column density than the HIP ($\log$N$_H>$21.8
[cm$^{-2}$])
requires $\log$U$>$2 (i.e 5 times larger U than
the HIP).
In fact the discrepancies (with significance $<2\sigma$) between
1.6 and 1.7 \AA, and near 10.6 \AA \ (see Fig. \ref{res_high}) hint
at the presence of such
material, as they are consistent with absorption by Fe {\sc
xxiv-xxvi}.

It is clear, then, that the gas tends to clump in two different
degrees of ionization, with the majority of the low ionized
material around the LIP. A third, very high ionization component
is hinted at by the data, but better S/N ratio data are needed to
establish its existence and constrain its physical properties.

\subsubsection{The Effects Induced by the Lack of Accurate Dielectronic
Recombination Rates \label{UTA}}

Our analysis of the LIP is based mostly on the Fe M-shell UTA, and
low temperature Dielectronic Recombination Rate 
(DRR) coefficients have not been calculated or measured yet for these ions.
Krongold et al. (2003) and Netzer et al. (2003) reported that the
models were unable to simultaneously reproduce the absorption by the UTA and
two lines of Si (Si{\sc x} at $\lambda 6.850$ and Si{\sc xi} at
$\lambda 6.775$). Netzer et al. (2003) proposed that this was an effect
introduced by the lack of these DDR. This means that, for a given
value of the ionization parameter, there
is a systematic shift in the ionization balance of Fe (toward higher charge
states) with respect to the rest of the species. Netzer (2004)
and  Kraemer, Ferland,
\& Gabel (2004) have estimated these rates. Their results show that there is
indeed an increase in the ionization parameter at which the Fe M-shell
ions peak, offering a solution for  the
mismatch between the UTA and Si lines in the spectrum of NGC 3783.

In the spectrum of NGC 985, besides the UTA, there are
no other significant absorption features that allow us to
constrain the ionization parameter of the LIP. Accepted at face
value, this means that we are underestimating the ionization
parameter (and temperature) of the LIP, because of the lack of Fe
M-shell DRR. However, from our previous analysis on NGC
3783 (Krongold et al. 2003), we estimate that the error cannot
exceed a factor of 2 in U (i.e. $\Delta \log U < 0.3$), a
factor consistent with the one obtained by Netzer et al. (2003)
and Kramer, Ferland, \& Gabel (2004). 
%This is in fact an upper limit
%since our code (PHASE, see krongold et al. 2003 for further details)
%derives the ionization balance from Cloudy (see Ferland et al. 1998),
%and Cloudy uses the
%approach by Ali et al. (1991) to estimate the Fe M-shell DRR (however,
%this estimates are not as good as the ones presented by Kraemer, Ferland
%and Gabel 2003, because they do not include the new data obtained by Gu
%2003). 
We have repeated our
analysis shifting up by a factor of 2 the ionization parameter of
the LIP in all elements except Fe. 
With the quality of the present
data, this distinction does not modify our results in any
significant way. In fact, estimates with this new approach of the
column density and ionization parameter of possible additional
absorbing components yield to limits fully consistent with those
presented in the former section, showing again that the gas clumps
in two, or maybe three, absorbing components.

While our main conclusions will not change due to this underestimation
of U, this effect should be kept in mind. Therefore, we have set the
positive error bar in the ionization parameter of the LIP to be
consistent with a factor of 2 change in U, in order
to reflect the underestimation introduced by the lack of low
temperature DRR for Fe.

Recent calculations by Gu et al. (2003), and measurements by Savin et
al. (2002a, 2002b, 2003) have produced more accurate values for Fe 
L-shell DRR.
At temperatures close to the one derived 
for the HIP $\sim 10^6$ K
($\sim 86$ eV) the differences between 
the old estimates and the new
ones are relatively small  (although these differences become 
larger at lower temperatures, i.e. $10^5$ K). Thus, although the new
calculations have not been incorporated into the photoionization
codes, we expect the uncertainties introduced into the HIP model by the
uncertainties of the 
Fe L-shell DRR currently used in the codes to be within
the error bars derived in our analysis. In fact, our model for this
component is able to fit the 
Fe L-shell lines, along
with K-shell lines arising from other elements, like Si, Mg, and Ne
(see Fig. \ref{spectra}). Although the significance of these narrow
lines is limited, it is clear that the ionization state of the gas is
consistent with the observed charge states. We note that the same
was found for the high S/N ratio spectrum of NGC 3783 (Krongold et al. 2003;
Netzer et al. 2003). Therefore, we do not expect a systematic
underestimation of U in the HIP.

\section{BeppoSAX DATA \label{sax_dat}}

To further study the nature of the ionized absorber, in particular
to constrain the absorber response to the changes in the continuum
of the central source, we retrieved and reprocessed the previously
unpublished NGC 985 data obtained by BeppoSAX (Boella et al. 1997;
Piro et al. 1997) on 1999 August 29 (see Table \ref{log}). The data
have a total
MECS exposure of 90 ks. We followed the standard reduction procedure
(Fiore et al. 1999) for the LECS, MECS, and PDS data. We extracted
LECS and MECS source spectra from circles of radius 8' and 4'
respectively, and the same extraction regions were used to extract
background spectra from the LECS and MECS ``blank fields.'' The
PDS instrument has no imaging capability and is a collimated
detector with a field of view (FOV) of 1$^o$.3 (FWHM). It makes
use of rocking collimators for background monitoring. Using the
ROSAT All Sky Survey, we searched the FOV of the PDS in the
direction to NGC 985, and did not find any other source in this
region capable of a significant contamination of the PDS spectrum
(Mrk 1044 is the closest bright X-ray source to NGC 985 in the
sky, at an angular distance of 1.2$^o$, but it was outside the FOV of
the PDS during the observation).

Modest variability was observed during the BeppoSAX
observation of NGC 985. In the 0.1-2.0 keV  band (LECS data),
source variations were detected up to a factor 1.45 on
a time-scale of about 50 ksec. Variability in the 2.0-10 keV range
(MECS data) was detected up to a factor of 1.3 in the same
time-scale. The quality of the data does not allow for a separation
in low and high state spectra, therefore we modeled the
full time integrated dataset.

%n Figure \ref{lc}c, we present the ratio of the 0.1-2.0 keV and
%.0-10 keV flux, there is a marginal indication of variation in
%he soft X-ray band with respect to the hard-Xray, as the spectra
%seems to gets softer and then harder, again in time scales of
%nearly 50 ksec. This behavior is suggestive of variability not
%only in the power law component of the continuum, but also in the
%soft excess. {\em Chandra} data further reveals variability of the soft
%component (see Table \ref{tabcont}).

\subsection{Spectral Analysis}

%As before, we used PHASE coupled to Sherpa to carry on the spectral
%analysis of the BeppoSAX low resolution data.

\subsubsection{Fitting the Continuum \label{sax_cont}}

\begin{figure}
\epsscale{1.0}
\plotone{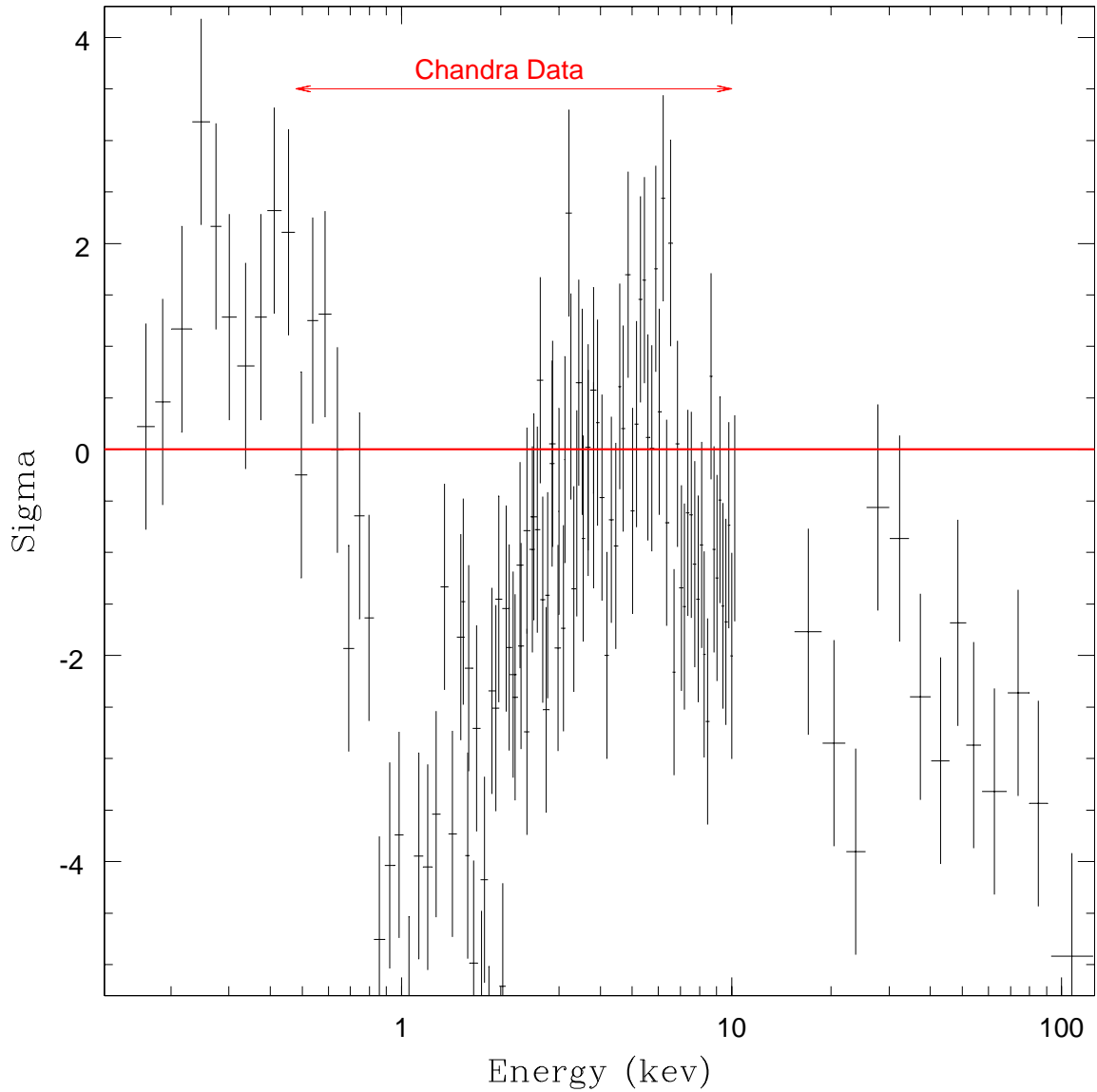} \caption[f12.eps]{Fit residuals for a
model including only a continuum power law constrained in the 3-8 keV
range and further attenuated by Galactic absorption,
over BeppoSAX data of NGC 985. {\it [See the electronic edition of the
Journal for a color version of this figure]}
\label{sax_res}}
\end{figure}

\begin{figure}
\plotone{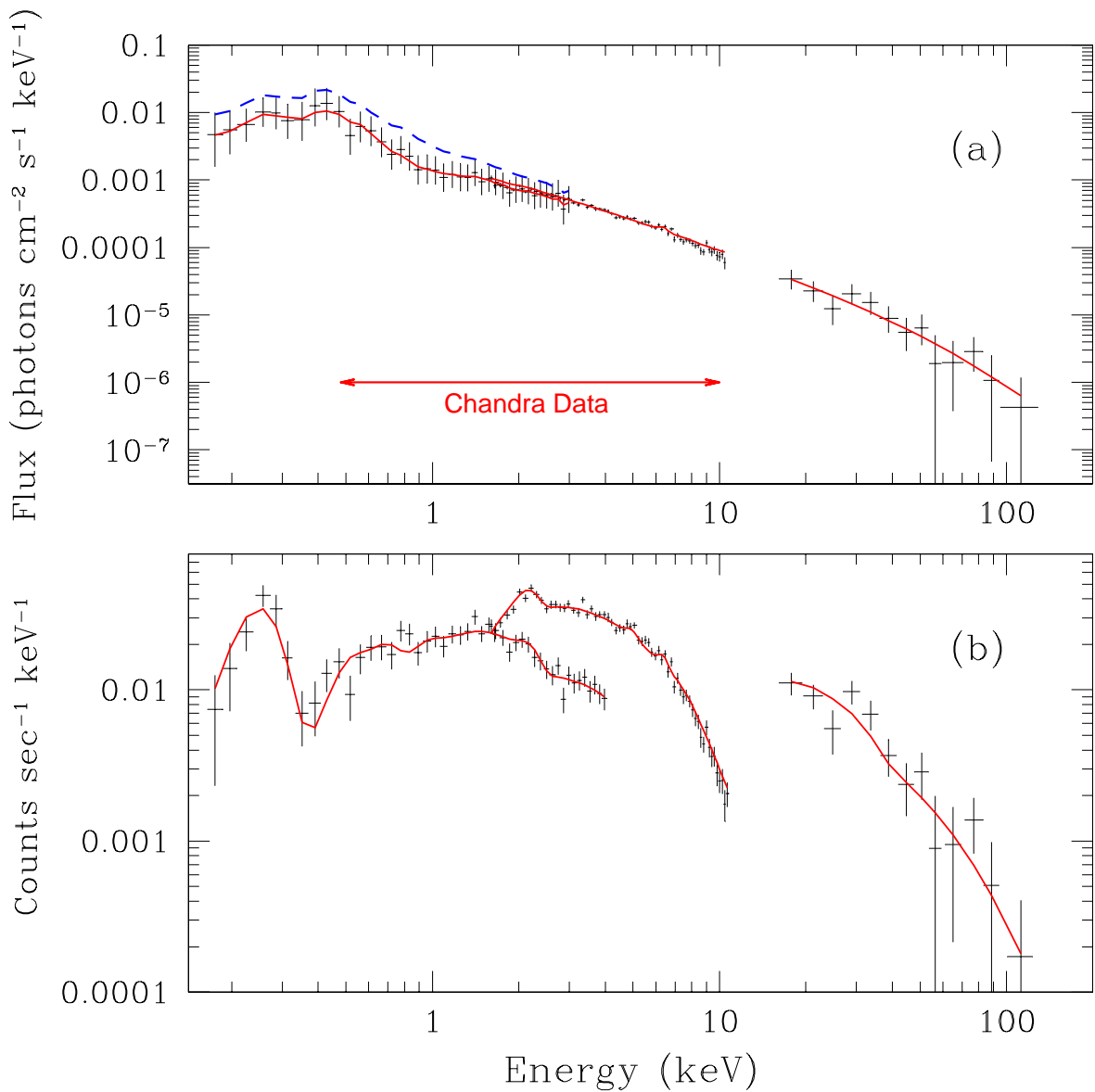} \caption[f13.eps]{BeppoSAX spectra of NGC
985, plotted against an ionized absorber model for comparison. (a)
Fluxed spectrum. The blue dashed
line shows the continuum further attenuated by Galactic absorption.
(b) Empirical Spectrum. {\it [See the electronic edition of the
Journal for a color version of this figure]}
\label{saxspec}}
\end{figure}

To model the broadband (0.1-100 keV) BeppoSAX continuum of NGC
985, we first followed the same approach used with the {\em
Chandra} data, constraining the intrinsic powerlaw in the 3-8 keV
range. The residuals of the extrapolation of this model to the
whole BeppoSAX band are presented in Figure \ref{sax_res}. The
presence of both an ionized absorber and a soft excess is evident,
together with that of a high energy cutoff at energies larger than
60 keV. We then refitted the data including an exponential cutoff
to model the continuum at high energies, as well as a blackbody
and an ionized absorber (see \S \ref{abs_sax}) to model the soft
excess and the absorption at low energies.\footnotemark\ Results of
our best continuum fits are shown in Table \ref{tabcont}. The
intrinsic continuum of the source further attenuated by Galactic
absorption is shown in Figure \ref{saxspec}a (dashed line).
\footnotetext{The inclusion of a cold reflector, often observed in
the high energy spectra of Seyfert galaxies (e.g. NGC 3783, Kaspi et
al. 2001, de Rosa et al. 2002), is not required
by the BeppoSAX data of NGC 985. We find a 2$\sigma$ upper limit
for the amount of reflection R$<0.7$, see Table \ref{tabcont}.}

%To model the power law component we considered only the 3-8 keV band
%(excluding the region of the Fe K$\alpha$ line),
%where no substantial effect from the absorber is expected. We included
%an exponential cutoff to fit the continuum at higher energies. We find
%only marginal evidence of a Compton reflection hump in the spectrum of
%NGC 985. We find a reflection
%factor consistent with 0 (R$<0.7$ within a significance level
%2$\sigma$). Such component, if present, would require a large
%inclination  angle
%($>80^o$), and therefore would have a weak
%contribution to the continumm below 8 keV, so it is unlikely to change
%the power law measurements presented here.
%Table  presents the values for the best continuum
%fit. The intrinsic continuum of the source, the blackbody soft excess
%component, and the continuum further attenuated by Galactic absorption
%are shown in Figure \ref{saxspec}a.

\begin{figure}
\plotone{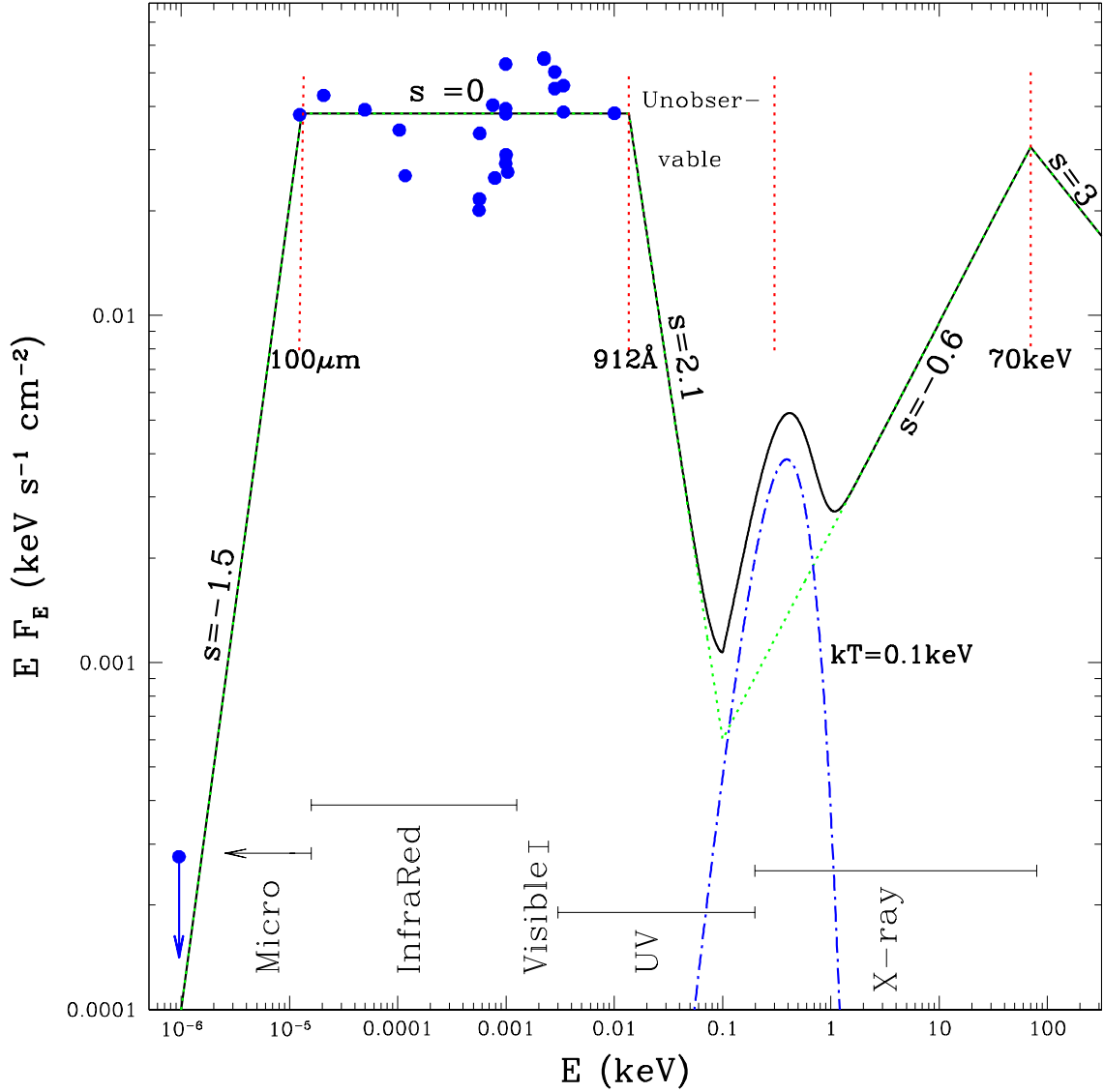} \caption[f14.eps]{Spectral Energy
Distribution used to model the ionized absorber observed in the
BeppoSAX spectrum. Filled circles (blue) mark the observed SED of
NGC 985 (obtained from NED). The plot shows the slope adopted in each
energy range. This slope relates to the photon index as
$\Gamma$=s+2. The model assumes
a low energy far UV cutoff at 0.1 keV and includes the
contribution of a blackbody component.{\it [See the electronic edition of the
Journal for a color version of this figure]} \label{sed_sax}}
\end{figure}

The BeppoSAX data indicate a flatter power law photon index
($\Delta \Gamma \approx 0.2\pm 0.07$) and a larger flux in both
the power law and the blackbody components (by a factor $\approx$
2.3) with respect to the Chandra data. These differences are reflected
in the SED used to fit the
ionized absorber in the BeppoSAX data (see Fig. \ref{sed_sax}).
%is
%similar to the one for the {\em Chandra} analysis (\S
%\ref{abs_chandra}),  but with a photon index $\Gamma=1.40$ above
%0.1 keV and a larger contribution from the blackbody, as indicated
%by the continuum fits presented here.

%The presence of the soft excess was more difficult to constrain,
%as it blends with the features imprinted by the ionized absorber.
%We used a blackbody component to model this excess. A larger
%column density for the ionized absorber would imply a larger
%normalization in the blackbody, to compensate for the deeper
%absorption. As will be explained in \S \ref{abs_sax}, we used the
%{\em Chandra} data to constrain the column density of the ionized
%absorber observed in the BeppoSAX data. This allowed us to
%estimate in a better way the soft excess component.

\subsubsection{Modeling the Ionized Absorber Present in the BeppoSAX
  Data  \label{abs_sax}}

Modeling of absorption features in low resolution data is extremely
difficult because the lack of detail does
not allow the determination of the exact ionization state of the gas and the
number of absorbing components. Because of this, we decided to
use the solution found for the high resolution {\em Chandra} data,
to constrain the
physical properties of the gas at low resolution. We assumed that
two components contribute to the absorption, as inferred from the
{\em Chandra} data, and fixed the outflow velocity and the turbulent
velocity of the absorbers to the values deduced in \S
\ref{abs_chandra} (these parameters are essentially unconstrained
when modeling low resolution data).

First, we left the ionization parameter and the column density of both
components free to vary while fitting the data. An acceptable fit
was obtained (reduced $\chi^2=1.063$ for 112 degrees of
freedom), with
column density values in both absorption components closely consistent
with the
values obtained modeling the {\em Chandra} data: we find column
density values of $\log$N$_H$=21.95$\pm.24$ [cm$^{-2}$]
($\Delta\log$N$_H$=0.14 [cm$^{-2}$])
for the HIP and $\log$N$_H$=21.38$\pm.26$ [cm$^{-2}$]
($\Delta\log$N$_H$=0.01 [cm$^{-2}$])
for the LIP. Hence, we next froze the column
densities to the values obtained with {\em Chandra}, and fit
the data letting only the
ionization parameters vary. The results are
reported in Table \ref{tab_abs_sax}, and are discussed in the
next section. Figure \ref{saxspec} shows the
data and the best fit model and Table \ref{ionic} shows the predicted column
densities for the dominant charge states.

\section{VARIABILITY ANALYSIS \label{vari}}

\subsection{Intensity and Spectral Changes in the Continuum}
A comparison of the 2002 {\em Chandra} and 1999 BeppoSAX
observations indicates a flux drop by a factor of 2.3 in the
0.1-10.0 keV range. Variability is observed in both the soft X-ray
(0.1-2.0 keV) and hard X-ray (2.0-10.0 keV) bands.  Since the hard
X-ray region is barely affected by the ionized absorber, the
change of flux is indicative of intrinsic source variations. Our
continua fits indicate a steepening in slope of $\Delta \Gamma
\approx 0.2\pm 0.07$. Any attempt to fit both datasets with a
single photon index value yields unsatisfactory results. For
instance, we find a $\chi^2=59.1$ for 50 degrees of freedom when
fitting (between 3 and 8 keV) the {\em Chandra} data with a value
of $\Gamma$ frozen to 1.4, this imply a $\Delta\chi^2=12.4$ with
respect to the best fitting model (an F-test gives a significance
larger than 99.9\% for this difference). Fitting the BeppoSAX data
with a photon index equal to 1.6 produces even worse results. This
difference could be indicating a real slope variation or could be
the result of cross-calibration effects between the two
instruments. We will follow the analysis assuming $\Gamma$=1.4 for
the BeppoSAX data, noticing that this difference does not have a
significant effect on our conclusions.

To investigate the significance of the variability of the soft
emission component,
% cannot be explained so
%straightforwardly, as it may be due to a change in the
%opacity of the absorber, variability of the excess, or both. To
%further study these possibilities we considered several scenarios
%consistent with only a change of opacity in the absorber. For this
%effect,
we attempted to fit the BeppoSAX data while freezing the
blackbody component used to parameterize the soft excess to the values
obtained for our best fitting {\em Chandra} model. Every model tried imposing
this constraint led to an underestimate of the continuum level below
1 keV, pointing to a genuine variation of the flux of the soft
component between the two
observations. We note from Table \ref{tabcont} that only a change in
the intensity of the emission is required by the data, as the best fit
solutions to both datasets are consistent with a blackbody temperature
of 0.1 keV.

\subsection{Changes in the Opacity of the Ionized Absorber}

The ionization degree of the ionized absorber along the line of
sight to NGC 985 has also changed between the {\em Chandra} and BeppoSAX
observations.  An F-test comparing the opacities deduced from the
{\em Chandra} best fitting model with those for the BeppoSAX best
fitting model shows that the difference is
statistically significant at the 99\%
confidence level. To better
quantify the changes in the ionized material, it is convenient to
introduce a different ionization parameter, U$_x$, defined through
the number of ionizing photons between 0.1 and 10.0 keV (Netzer
1996). This definition makes the changes in the ionization parameter
(now relevant only to the
X-ray species) directly proportional to the change in the
X-ray flux, as opposed to bolometric flux changes, which are not
directly observable. Thus, it is preferable in studies of the
effects of flux variability over ionized gas in the X-ray regime.
The two different definitions of the ionization parameter scale as
$\log$U$_x$=$\log$U - 2.43 for the {\em Chandra} SED, and
$\log$U$_x$=$\log$U - 2.08 for the BeppoSAX SED.

The best fit to the {\em Chandra} data gives a value of
$\log$U$_x$=-2.55$\pm0.09$ for
the LIP. If the gas is in photoionization
equilibrium and is not physically confined, a change of flux by a
factor 2.3 would induce a change
in the X-ray ionization parameter by the same factor, yielding a value
of $\log$U$_x$=-2.20. The best fit value of $\log$U$_x$ for the
BeppoSAX data is -2.30$\pm0.10$, consistent with the predicted value.
%  Given the uncertainties obtained for U
%and the flux, we notice that these two values are consistent with each
%other. In fact, modelling of the SAX data constraining $\log$U$_x$ to
%the expected value of -2.20 yields a $\chi^2=1.090$ for 115 degrees of
%freedom.  Following an f-test with a difference of 1 interesting
%parameter (the ionization parameter of the LIP) confirms that the best
%BeppoSAX model for the LIP and this new model are not statistically
%different.
We conclude that the LIP is responding to the changes in the
continuum and thus is probably both photoionized and in equilibrium.

The value of $\log$U$_x$ for the HIP in the {\em Chandra} data model is
-1.09$\pm0.10$. If the absorber is in equilibrium, the flux change observed
between observations implies a value of $\log$U$_x$=-.73 for the HIP
present in the BeppoSAX spectrum.  The value obtained in our best
model for the BeppoSAX data is $\log$U$_x$=-1.25$\pm0.13$. These two
values of
the ionization parameter are clearly inconsistent with each
other.
Thus, this component has indeed
changed ionization state between the two observations, but does not seem to be
following the changes on the continuum. On the contrary, the degree of
ionization of this gas seems to have increased while the flux
decreased.

Such behavior has several possible explanations. (1) The gas is out
of equilibrium: the photoionization equilibrium time scale is
inversely proportional to the density in the gas during both
increasing and decreasing flux intensity phases (e.g. Nicastro et
al. 1999) . If the density is low enough, this timescale can be
larger than the variability timescale of the central source, and
the medium does not have time to adapt to the changes in the
continuum. In such conditions, delays in gas responses may lead to
scenarios where the ionization degree of the gas decreases while
the flux increases. Such cases have been already found in other
ionized absorbers around AGN, for instance NGC 4051 (Nicastro et
al. 1999). (2) The gas is not in photoionization equilibrium
because another mechanism, collisional ionization (due, for instance,
to shock heated gas) dominates. In
this case, obviously, changes in the gas opacity are essentially
unrelated to changes in the source flux. (3) The gas is indeed in
photoionization equilibrium, and has undergone a significant
change 
in its density. This is possible if the gas is
pressure confined by an external medium. (4) Finally, it is possible
that the absorbing gas is not the same during the two
observations. X-ray absorbers can change independently of flux
variations (see George et al. 1998) because of transverse motion of
the material or because of condensation and dissipation of clouds out
of a hot medium (Krolik \& Kriss 2001). 
In the
following sections we will explore further the validity of these
different scenarios.

\section{ABSORBER PHYSICS DISCUSSION \label {disc}}

The ionized absorber present in NGC 985 is
suggestive of a simple picture, being constrained by two strong
absorption components. The same situation was found for
the ionized absorbers in IRAS 13349+2438 (Sako et al.  2001)
and NGC 3783 (Krongold et al. 2003).
In all three objects, an Fe {\sc vii-xiii} UTA and several
\ovii \ lines are the signature of the cooler phase, and the \oviii \
and Fe {\sc xvii-xxii} are the main features of the hotter phase, with
high ionization iron dominating the spectrum below 15 \AA. These
ionized absorbers are inconsistent with substantial amounts of
material with intermediate values of the ionization
parameter. Furthermore, the ionization degree of the LIP is consistent
with a cooler medium producing significant amounts of O {\sc vi} and
other lower ionization species, which should also imprint their
signature in the UV band (see also \S \ref{uv}). In the case of
NGC 3783, a third
very high ionized component has also been detected (Netzer et
al. 2003). In NGC 985 the
presence of such component is hinted by the data, although it cannot
be well constrained (see Fig. \ref{res_high}).

The similar characteristics in the absorbers of these objects
suggest that this an intrinsic property related to the structure
of the nuclear environment. In the following sections, a possible
scenario regarding the nature of these absorbers is discussed.

\subsection{The UV X-ray Connection of the Absorbers \label{uv}}

NGC 985 has been observed with HST-STIS and FUSE.  Arav (2002)
modeled these  data, and showed 
that, once saturation is accounted for, 
the component outflowing at 404 \kms\ (i.e. component 4)
produces absorption in the X-ray band.

In Table \ref{uvtab} we present the O {\sc vi}, N {\sc v}, C {\sc iv},
and H {\sc i} column densities inferred from the LIP of our {\em
Chandra} best fitting model. It is evident that this absorbing
component would produce absorption 
in the UV band. Arav (2002) measured column densities of
$\log N_{Nv}=14.2$ [cm$^{-2}$], and an $\log N_{H_I}=14.7$ [cm$^{-2}$]. 
These values are smaller than the ones derived from our model by
factors 29 for N {\sc v} and 5 for H {\sc i} (see Table \ref{uvtab}).
Finding consistent column densities for the UV and X-ray absorbers has
been historically difficult as has been discussed by Krongold et
al. (2003). In the case of NGC 985,
a few reasons can be responsible for the discrepancy:
(1) The predicted column densities inferred by any model in the UV
strongly depend on the exact shape of the SED assumed in the analysis
(Kaspi et al. 2001;  Steenbrugge et al. 2003). Thus, the
overestimation in the model could be an effect induced by our chosen SED
(Fig. \ref{fsed}a). In fact, the
derived factors reduce to 15 for N {\sc v} and 4 for H {\sc i} when
considering the SED presented in Figure \ref{fsed}b (see Table
\ref{uvtab}). The reason
for the decrease in ionic column density in this SED is
that it has a larger number of photons in
the extreme UV, the relevant region for the abundances of the ions
absorbing in the UV (note
the small decrease in H {\sc i}, since the photon flux assumed in both
SEDs at 912 \AA\ is the same). However, the intrinsic SED in this
region cannot be constrained
due to Galactic absorption. 
(2) In addition to SED effects, the fact that
the observations are not simultaneous  should be kept in mind. It is
possible that the absorber
has changed in opacity because the gas responds to flux
variations (or because the absorbing gas is not the same at the two
different epochs). For instance, the ionization parameter for the LIP
in the BeppoSAX model is larger than the one in the {\em Chandra}
model. Then,  the difference in N {\sc v} column
density between Arav's measurements and our BeppoSAX best fit model  
is only 13, if the SED presented in Fig. \ref{sed_sax} is used (this
SED is similar in the extreme UV to the one presented in Figure
\ref{fsed}a for the {\em Chandra} analysis, with a small flux in this
region).  
%If instead, an SED with a
%larger extreme UV flux is considered the discrepancies would reduce
%even more. 
Therefore, the different columns can reflect only the
ionization state of the gas. 
%Thus, the
%difference can be attributed
%to (1) the exact shape of the extreme UV
%SEDs and (2) possible
%changes in the opacity of the gas due to flux variations (the
%X-ray and UV observations are not simultaneous), and 
(3) Finally, it is possible that other UV components (besides
component 4) contribute to the absorption observed in the X-rays. 

Quantifying these differences would require simultaneous UV and X-ray
observations with
high resolution and high S/N ratio. However, we note that NGC 985, along
with NGC 5548 (Kaastra et al. 2002; Arav et al. 2003; Steenbrugge et
al. 2003), NGC 3783 (Kaspi et al. 2001; Gabel et al. 2003; Krongold et
al. 2003; Netzer et al. 2003), and NGC 7469 (Blustin et al. 2003;
Kriss et al. 2003) are good examples of the UV and X-ray absorbing connection.

\subsection{Density and Location of the Absorbing Gas \label{loc}}

Our analysis of the LIP indicates that this gas may be in
photoionization equilibrium and responding consistently to the
changes in the continuum. Using the elapsed time between the high
state and low state observations (1029 days) as an upper limit to
the recombination time, a lower limit on the density of the
recombining gas and an upper limit on the distance from the
central source can be set for this gas. For this calculation, we
used the photoionization equilibrium temperature, the flux in the
0.1-10 keV range derived from our fit to
the {\em Chandra} observation, the ionization parameter U$_x$, and
the recombination rate for \ovii \  (Shull \& van Steenberg 1982)
since 
it is the dominant charge state. We estimate a lower limit for the
electron density of the
LIP of n$_e$(LIP)$>$610 cm$^{-3}$. This translates into an upper
limit of R(LIP)$<$ 21.6 pc for the distance between the central
source and the ionized gas, and a maximum thickness
$\Delta$R(LIP)$<$ 1.3 pc for the gas.

%We cannot assume that the HIP is in equilibrium since it is not
%following the continuum changes. However, under the assumption that
%the gas did not recombine between the observations (i.e. the gas out of
%photoionization equilibrium) it is possible to
%set now, an upper limit on the density and a lower limit on the
%distance to the central source. We notice that this assumption is very
%weak, since the source presents variability up to a factor 1.5 in
%timescales of less than one day (see \S \ref{sax_dat}). Therefore it
%is difficult to think that a change of flux by a factor of 2.3 is a
%unique event in the 1029 days elapsed between observations. Thus,
%if the HIP is out of equilibrium, it could be due to a variability
%event that took place close to the time of the {\em Chandra} observation,
%making our estimations meaningless. Keeping this caveat in mind, and
%using the recombination rate from \oviii \ (Shull \& van Steenberg 1982), we
%obtain n$_e$(HIP)$<$3300 cm$^{-3}$, R(HIP)$>$ 0.1 pc, and $\Delta$R(HIP)$>$
%0.6 pc.

%the ratio $\Delta R/R < 1$, where $R$ is the
%distance from the HIP gas to the central source and $\Delta R$ the
%radial thickness of the absorber, it is possible to obtain a lower
%limit for $R$ and an upper limit for the electron density. This is
%because  $\Delta R/R =4\pi R$N$_H$U$_x$c/Q$_x$. Under this condition
%R(HIP)$<$4.7 pc and n$_e$(HIP)$>$4.5$\times 10^2$ cm$^{-3}$.

\subsection{Thermal Instabilities and the Absorbing Components  \label{term}}

Several models have been proposed to explain the nature of the
ionized absorbers around AGN. The different scenarios include a
continuous range of ionization parameters (Krolik \& Kriss 2001),
different clouds in different regions (including for instance
different UV and X-ray absorbing regions), or
similar locations for the absorbing winds (Krongold et al. 2003).

The data presented here suggest that a continuous range of U 
does not adequately 
describe the observed spectra (see
discussion in \S \ref{absfits}). The presence of different clouds
in different regions seems unlikely (at least for the objects
discussed here), 
not only because of the similar characteristics of the
absorbing components in different AGNs, but 
also because
pressure balance is observed (to within the errors) between the components (see
\ref{2phase}).

An intriguing possibility is that the different components form because of
thermal instabilities in the thermal photoionization equilibrium
S-curve (Krolik, Mckee, \& Tarter
1982). This curve marks the points of thermal equilibrium in the
T, U/T plane, where T is the equilibrium temperature of the gas,
and U/T is a quantity inversely proportional to the pressure of
the gas (see Fig.  \ref{xi_chandra}). The wiggles in the S-curve are
due to changing heating and cooling
rates, as different ions become dominant. The shape of this curve is
also affected by the ionizing SED and the metallicity of the absorbing
gas (e.g. Komossa \& Fink 1997). The high temperature branch lies
where only Compton heating is important. Gas in regions of the curve
with negative derivative is unstable because any isobaric
perturbation will be amplified, leading to net cooling or heating.
Then, the different
components observed form simply, when the gas is driven to the
stable (positive slope) branches of the curve, while tending toward
equilibrium.

%A deeper
%analysis of these scenario is in progress.

\subsection{Pressure Balance between the Absorbing Components \label{2phase}}

Krongold et al. (2003) reported that the two components in
the ionized absorber of NGC 3783 are in pressure balance
to within 7\%,
which suggested that the absorption could arise from two phases of the
same medium. As we shall point here, this also seems to be the
situation for the ionized absorber of NGC 985.

\begin{figure}
\plottwo{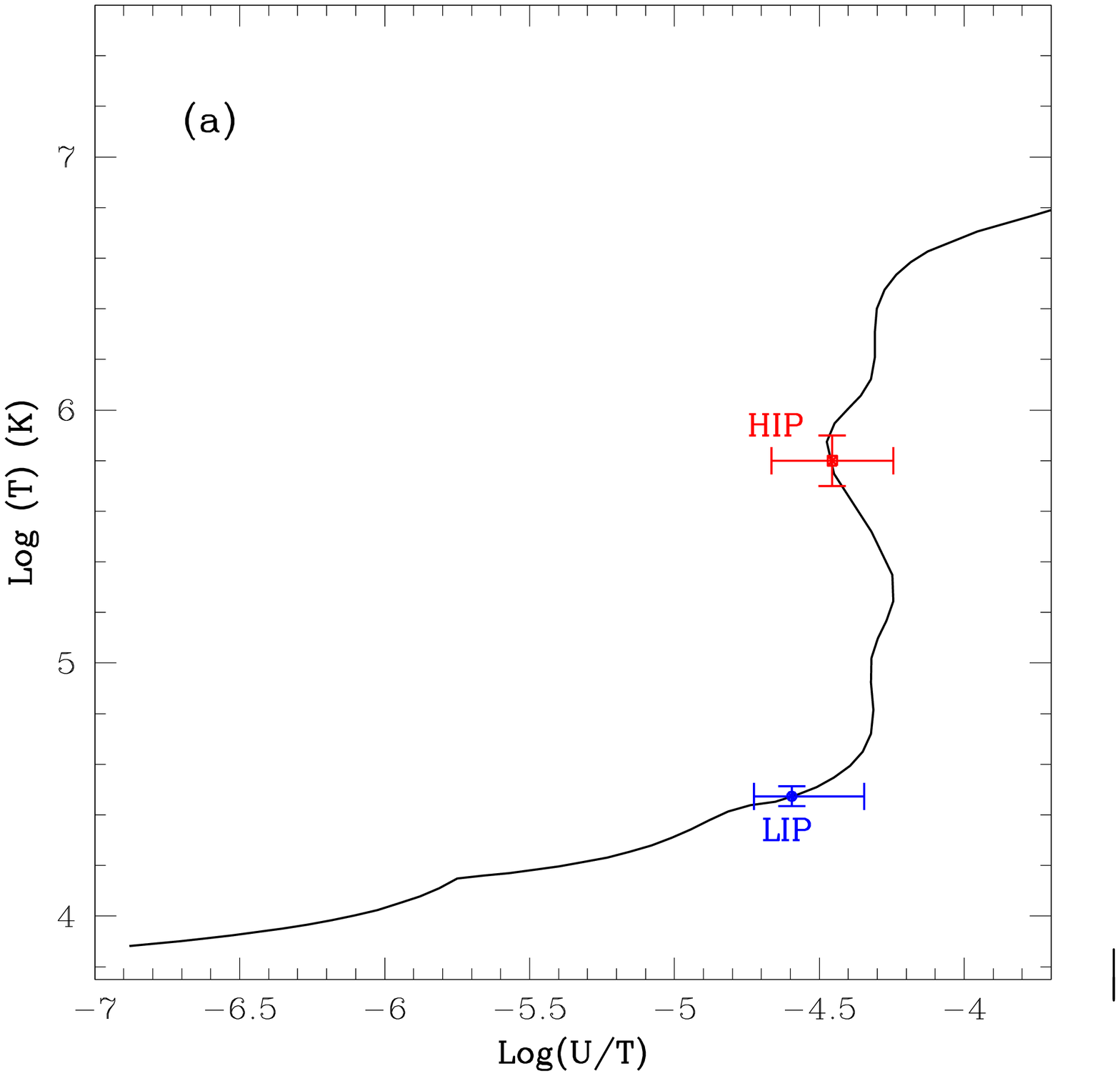}{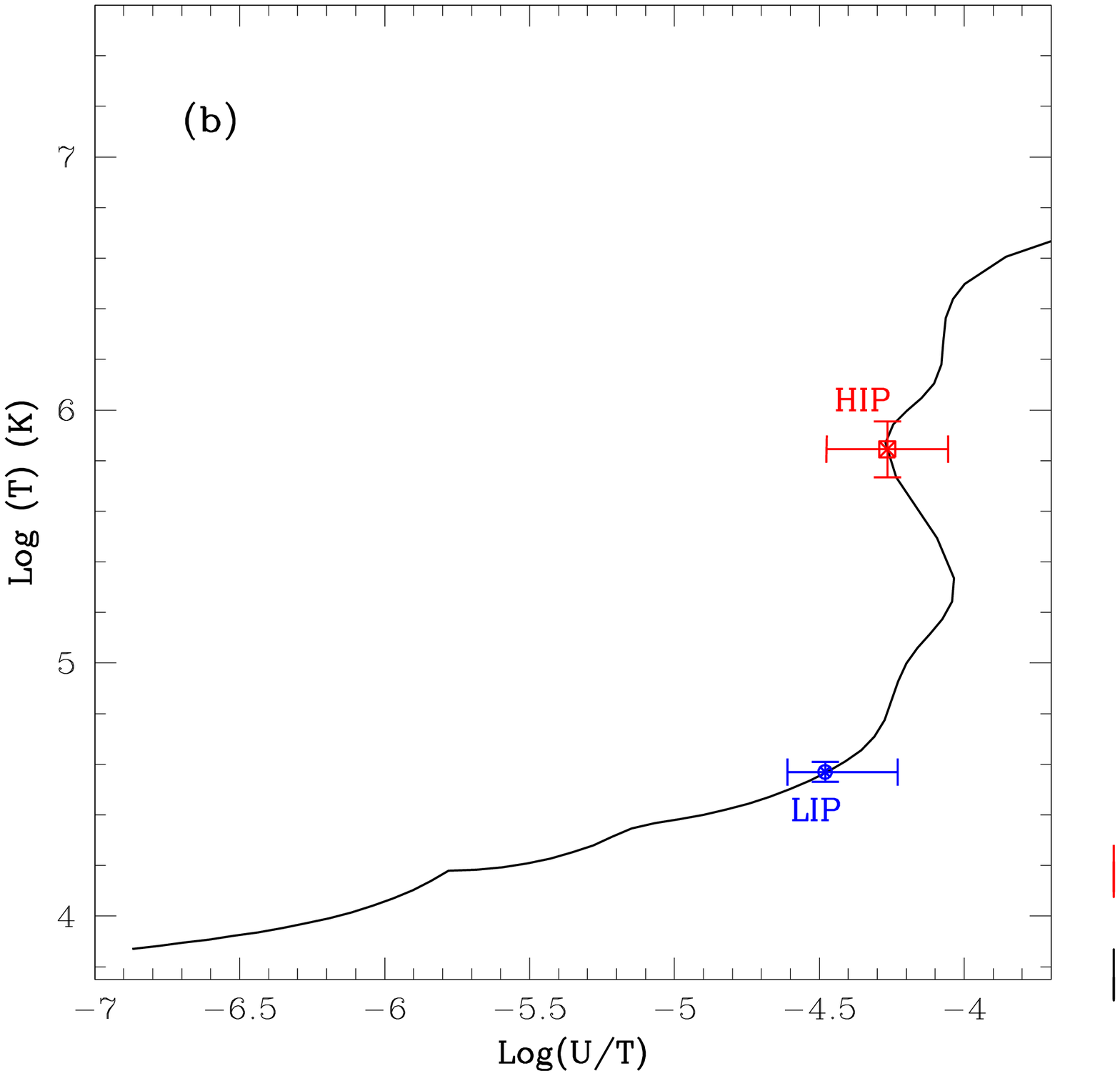}
\caption[f15.eps]{Curve of thermal
stability for the two different SEDs used in the analysis of
{\em Chandra} data, as described in the text. (a) Far UV cutoff at 0.1
keV plus blackbody (Fig. \ref{fsed}a). (b) Far UV cutoff at 1 keV
(Fig. \ref{fsed}b). The positive error bar of the LIP
represents the maximum effect in U induced by the lack of low
temperature DRR. The HIP and the LIP are
consistent with pressure equilibrium. {\it [See the electronic edition of the
Journal for a color version of this figure]} \label{xi_chandra}}
\end{figure}

As can be
observed in Figure \ref{xi_chandra}a, the HIP and the LIP in the
best fit model of the NGC 985 {\em Chandra} data have very
different equilibrium temperatures but lie close to each other in
the U/T axis. As mentioned before, with the adopted definition of
the ionization parameter (U), this ratio is inversely proportional
to P, the gas pressure. Hence, assuming that
the HIP and the LIP have the same location, the gas pressure between
them is indistinguishable within 25\%, (the
pressure of the phases is consistent with a single value at a
level well within 2$\sigma$).

We stress again the fact that the  absorbing material
does not span all the points in the
stable branches of the S-curve, but rather it is concentrated in
separate phases (see \S \ref{absfits}). Therefore, although the effect of
thermal instabilities may be partially responsible for the formation
of the phases (\S \ref{term}), this cannot alone explain
the observations. One
possibility is that these phases form as the gas tends to go to
the stable branches of the thermal equilibrium curve, but keeping
pressure balance between them (see below).

There are two special considerations regarding the validity of
the pressure equilibrium result that have to be discussed at this point:

\noindent (1) The shape of the SED: Kaspi et al. (2001) and
Steenbrugge et al. (2003) have shown that the shape of the SED in
the far UV (the unobservable region of the spectrum) has a
negligible effect on the charge states absorbing in the X-ray
region. However, the curve of thermal equilibrium  does depend on
the shape of the SED at these wavelengths. Thus, it is possible
that SEDs with a UV-X ray break different to the one assumed here would
result in two phases lying out of pressure balance. In other words,
our result could be reflecting specific characteristics of our
chosen SED. To further investigate this possibility, we have
repeated our entire analysis with the SED presented in Figure
\ref{fsed}b (with the UV-X ray break at 1 keV instead of 0.1 keV).
This choice gives the maximum effect introduced in our results
by SED uncertainties since the break cannot be at larger energies
because then, the observed spectrum would be directly affected. We find that
pressure equilibrium is found regardless of the assumed shape of the
far UV SED (see Figure \ref{xi_chandra}b).

\noindent (2) The analysis of the LIP is based mostly on the UTA:
As discussed in \S \ref{UTA}, the lack of reliable 
DRR for low Fe charge states may be producing an underestimate of the
ionization parameter of the LIP. Even if this effect is real, it
does not modify the analysis presented here in any way. More
accurate DR rates would shift the position of the LIP to higher
values of both pressure and temperature in the S-curve ($\Delta
\log U < +0.3$, see \S \ref{UTA}). However, the shift is
insufficient to move the LIP out of pressure equilibrium with the
HIP. On the contrary, such a change would make the pressure of
both components even more closely matched (Figure
\ref{xi_chandra}).

\subsection{Pressure Confinement of the Absorbing Components \label{2phase2}}

\begin{figure}
\epsscale{.5} \plotone{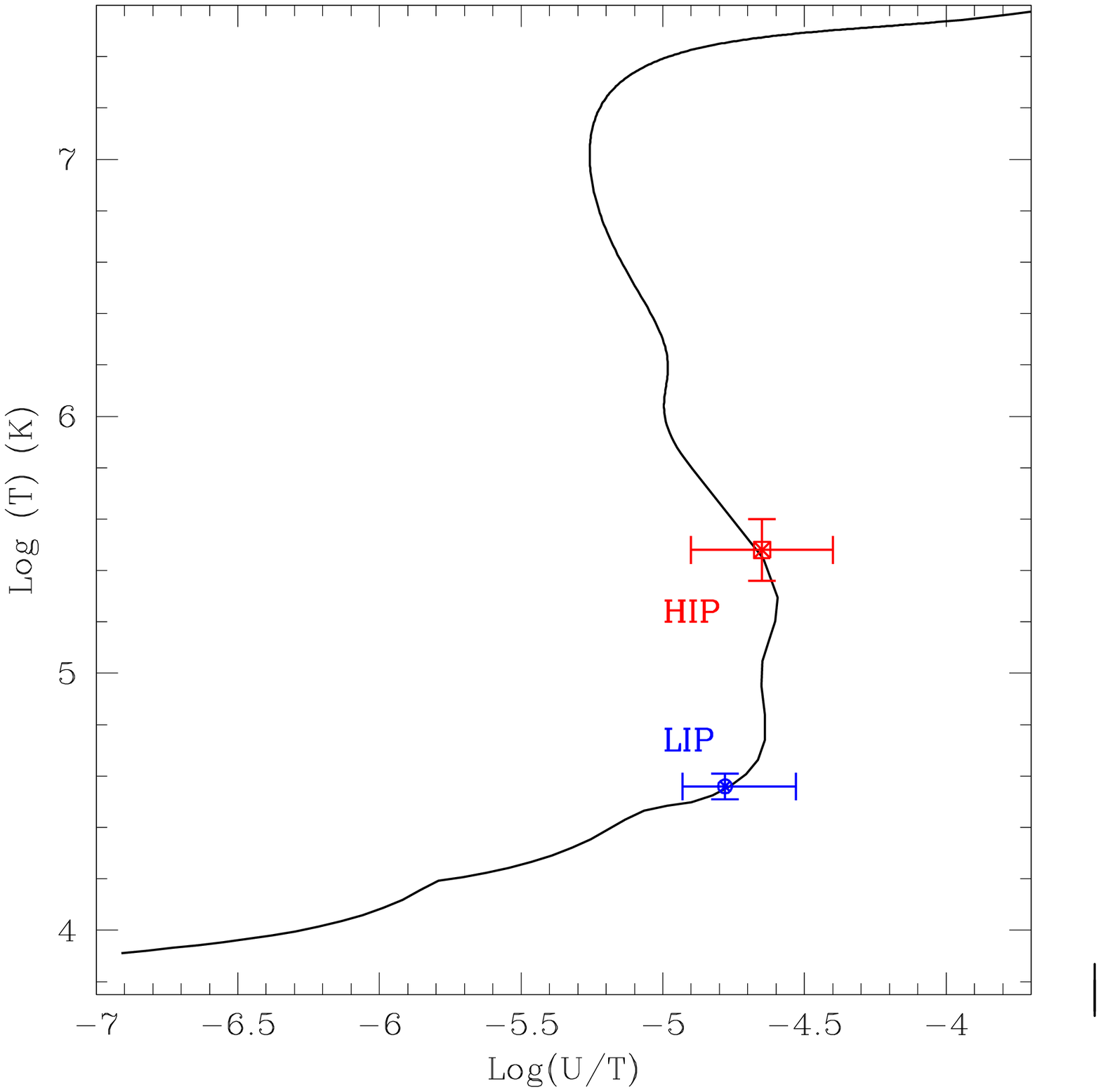} \caption[f16.eps]{Curve of
thermal stability for the SED used in the BeppoSAX analysis, as
described in the text. The positive error bar of the LIP
represents the maximum effect in U induced by the lack of low
temperature DRR. Pressure balance between the LIP and the HIP
appears to be present in the BeppoSAX data, even though the ionization degree
of the gas is different than the one obtained for {\em Chandra} data
(see text for details). {\it [See the electronic edition of the
Journal for a color version of this figure]} \label{xi_sax}}
\end{figure}

A striking result is that pressure equilibrium between the
two phases is also found in the BeppoSAX observation of NGC
985. This can be
observed in Figure
\ref{xi_sax}, where the S-curve for the SED used in the
BeppoSAX analysis is shown, and the positions of the HIP and the LIP in
this model are also marked. Although it is not clear what drives the
change of ionization in the HIP, both phases again align close to a
single value of $\log$U/T, as
would be expected if one of the two components confines the other one
(or as if both components are pressure confined by a third phase).

%(assuming the gas pressure dominates the total pressure).

There are two necessary conditions for this to happen: (1) The
hydrodynamic timescale (t$_H$) in the confined medium has to be shorter
than the timescale in which the gas reacts to the changes of the
continuum.
%Otherwise it
%would be easy to find the components out of pressure balance.
The sound velocity $v_s$ for media at temperature \simlt $10^6$ K
is \simlt 100 \kms, and the hydrodynamic timescale can be
estimated as $t_H \geq\Delta R/v_s$, where $\Delta R$ is the scale
size of the medium. For a single gas cloud with
 conditions typical of ionized
absorbers ($10^4$ K$<$T$<10^6$ K, and N$_H>$
few 10$^{21}$ cm$^{-2}$), the hydrodynamic
timescale will always be larger than the photoionization
timescale, which prevents the gas pressure from responding to the
change in the incident continuum quickly enough to remain in pressure equilibrium. This is true for any density inferred using the
ionization parameters from our model, assuming that the gas is
located anywhere within 10 parsecs of the central source. Thus, an
important conclusion is that the phases can only be in true
pressure balance if the confined medium is composed by a large
number of small clouds ($\sim 10^4$ clouds), so that the hydrodynamic
time scale
can be shorter than the photoionization timescale. The total
column density of the absorber would result from the individual
contributions of the clouds. 
%Interestingly, a scenario where the
%absorber is composed by many small clouds has already been
%suggested for the UV absorbers on NGC 3783 (Gabel et al. 2003).

(2) The photoionization equilibrium timescale of the media has to
be shorter than the variability timescale of the central source,
in order for the absorbing material to be close to
photoionization equilibrium.
This condition is assumed in our analysis. Clearly, if the gas is out
of photoionization equilibrium, then the ionization parameters and
temperatures derived here are not meaningful, and the absorbing gas
is not even constrained to lie on the S-curve.  We note
that photoionization equilibrium of the confining medium
is not required for pressure confinement (see \S
\ref{3erd}).

We will follow the discussion assuming that these two conditions are
indeed satisfied (but see \ref{nobal}). In such a
scenario, the confining medium would be responding to the changes 
in the continuum, and moving freely on the S-curve. The confined medium, on
the other hand, would not be able to move freely in the S-curve,
rather it would have to follow at all times
the position of the confining medium in the U/T axis. This means
that the confined medium would have to adapt to the ionizing continuum with a
constrained value of the pressure at all times. Thus, to reach the
photoionization equilibrium temperature, the confined phase would have to
expand or contract to keep pressure balance with the confining
medium.
%Assuming that the density in
%both phases is high enough, both could be in photoionization
%equilibrium.
Then, the ionization parameter (and thus, the opacity) of the confined
phase would be determined not only by the change in flux of the central
source (as in an isolated system), but also by the change in
density due to the change in volume.

%Considering the observed
%change in ionization state of the gas between the two
%observations, this implies a few scenarios for the physical
%conditions of the media that will be discussed in the following
%sections.

\subsubsection{Pressure Confinement of the HIP by the LIP}

This scenario can easily explain
the observed behavior of the HIP between the BeppoSAX and {\em
Chandra} observations: as
the flux drops, the LIP gets colder and its pressure drops. Hence,
the HIP would have to expand to decrease its pressure. The drop in
density of the HIP would increase its ionization parameter and
temperature (even though the flux is dropping), until
photoionization equilibrium is reached.
%The ratio of the density
%for the HIP in the low and high state of the source is given by
%the relation n$_L$/n$_H$=U$_H$/U$_L\times$F$_L$/F$_H$, where F is
%the flux, n the number density, and L and H stand for the high and
%low state respectively. Using the change in flux of 2.3 between
%the observations in the X-ray band, and U$_x$, this would require
%a drop in the density of the HIP by a factor of 3.3.

However, an inconsistency for this picture is that the column
density of the LIP is smaller than that of the HIP and its density
is larger (if the phases are at the same distance, the difference
in U between them should be mostly an effect of the different
densities). This would require the HIP to be more extended than
the LIP, and at the same time, the HIP would have to be made up of
several small clouds immersed in the LIP. We find it extremely unlikely
that these two conditions can be satisfied simultaneously.

\subsubsection{Pressure Confinement of the LIP by the HIP}

A second scenario is that the HIP is confining the LIP. However,
the HIP would have to be out of photoionization equilibrium, in
order to explain the changes in ionization parameter.
This picture is not possible, since as explained before, if this
phase is not in photoionization equilibrium, then its position on
the T-U/T plane is unknown. Thus, the alignment on the S-curve would
have to be considered just a bias induced by our analysis
(see \S \ref{nobal}).

\subsubsection{Pressure Confinement of the HIP and the LIP by a Third
Component \label{3erd}}

A third scenario is that a more ionized component is confining both the
HIP and the LIP. Being more ionized, this component would require
a density lower than that of the HIP (which is already 30 times
smaller than the
density of the LIP). In this case we have two 
possibilities. The first is that despite the low density, this component still 
has
time to adapt to the changes in the continuum, and therefore is in
photoionization equilibrium. The second is that this is not the
case. In either scenario, the HIP and the LIP would have to follow
the position of this hypothesized phase in the U/T axis.
In the photoionization equilibrium situation the third component has to be
located on the curve of thermal stability. In the non-photoionization
equilibrium case, the component could
be located off the curve, as long as the pressure imposed by this
phase on the other two is compatible with the region of the S
curve where multiple phases can survive, otherwise the phases
would dissolve.

This scenario is appealing because the presence of this
phase could easily explain the observed change of ionization
parameter of both the LIP and the HIP, between the BeppoSAX
and {\em Chandra} observations, as we show below. In addition, the
data hints
the presence of more ionized gas (see Fig. \ref{res_high}).
In NGC 3783, the presence of
such absorber has already been established, and found
to be in pressure balance with the less ionized components (Netzer
et al. 2003), which further suggests that this could be the case
in NGC 985. Evidently, much better data and time evolving
photoionization models (e.g. Nicastro et al.
1999) are needed to explore further this scenario.

If a third confining component is the
correct picture, then the HIP and the LIP have to be close to
photoionization equilibrium, but may not respond ``as expected''
to the flux variations because their pressure is fixed to the
pressure of the hottest phase. If this hot phase does not
change its position on the (U/T,T) plane dramatically in response to
continuum changes (because it is out of photoionization
equilibrium or because the gas lies on a steep part of the S-curve),
then the HIP and/or the LIP may appear as
if they were not following (or not even responding to) the continuum.

%This could explain
%why, in NGC 3783, no variation was observed by Behar et al. (2003)
%and Netzer et al. (2003). Then, the density and location of these
%absorbers could be completely different than those inferred by
%these authors. In particular, the density could be larger and the
%gas could be much closer to the central source, as suggested by
%the UV absorbers (Gabel et al. 2003). This idea seems to be
%supported by the findings of Reeves et al. (2003), who reported
%variation of the most highly ionized absorption lines in the X-ray
%spectrum of this object.

\subsection{The Absorbing Components Are Not in Pressure Balance \label{nobal}}

From the last sections, it is clear that a scenario of
pressure-confining is appealing because, due to the regions of thermal
stability in the S-curve, it can explain the different absorption
components observed in the spectra of Seyfert galaxies. However,
such scenario also imposes strong physical conditions on the
media that could be difficult to satisfy. There is not enough
evidence currently to rule out or confirm such 
a scenario and other
possibilities have to be considered.

% but the idea
%should be tested further, and .

If the pressure confinement is not real, why do the phases align
close to a single value of U/T? The answer to this question may
rely on the shape of the S-curve. For several SEDs, the S-curve
can be extremely steep in a large range of
temperatures (between 10$^4$ to 10$^6$ K) where photoionized gas
absorbs in the X-ray range (Krolik \& Kriss 2001). 
Large gradients of temperature (factors 10-50)
can 
then arise from small changes of the gas pressure. This is clearly
illustrated in Figures \ref{xi_chandra} and \ref{xi_sax}.
Thus, if several absorbing components were
in photoionization equilibrium and located in the ``vertical
region'' of the S-curve, but were completely disconnected from
each other, they would appear as if they were in pressure balance.
While this is certainly plausible, we note that
there are also SEDs that do not show such a pronounced vertical
branch (Komossa  \& Fink 1997). In those cases, it is possible to
disentangle true
pressure balance from apparent pressure balance (see for example
the S-curve obtained for the SED derived by Krongold et al. 2003
in their analysis of NGC 3783, their Fig. 16). Thus, discriminating
true pressure balance from apparent pressure balance
strongly depends on (1) the quality of the X-ray spectrum to constrain
better the position of the phases in
the (T,U/T) plane, and (2) the true shape of the SED, which 
is, however, always uncertain.

Evidently, if the gas is out of photoionization equilibrium, the
curves of thermal
stability discussed here are not relevant to describing the physical
state of the gas. Then, we would not know if the phases
are or are not in pressure balance. We note that the excellent fits
obtained for the high
quality data of NGC 3783 (Krongold et al. 2003, Netzer et al. 2003)
suggest that the gas is at least close to photoionization
equilibrium. Time evolving photoionization models applied 
to high quality data are the best way to determine if the data are in
photoionization equilibrium, and to further study possible scenarios of
pressure confinement out of equilibrium.

Additional complexity is present because the X-ray detectors are
most sensitive to absorption features produced by gas in exactly
the vertical region of the S-curve. So, the detection of
components only in the vertical branch (i.e. in apparent pressure
equilibrium) could be the reflection of an observational bias.
%due to
%the limited S/N of the data (simply because the detectors do not
%see the cooler gas, that lies on the horizontal branch of the
%S-curve). According to Figure \ref{cont}a, there is no gas with
%$\log$N$_H>21$ [cm$^{-2}$] and ionization parameter smaller than that of the
%LIP ($\log$U$<-.12$), or would be detected in the spectrum.
To explore this possibility we have simulated a
series of {\em Chandra} observations using the
same exposure time and the same best
continuum model for the NGC 985 observation (see Table \ref{tabcont}).
We have simulated data with two different absorption scenarios:

\noindent (1) The first dataset was produced using the HIP plus gas less
ionized than the LIP (i.e. gas located to the left of the LIP in the
horizontal branch of
the S-curve with $\log U$=-0.8 and
$\log$N$_H$=21.5 [cm$^{-2}$]). We modeled these data with PHASE and found that
the presence of colder gas was
easily detected and constrained.

\begin{figure}
\epsscale{1}
\plottwo{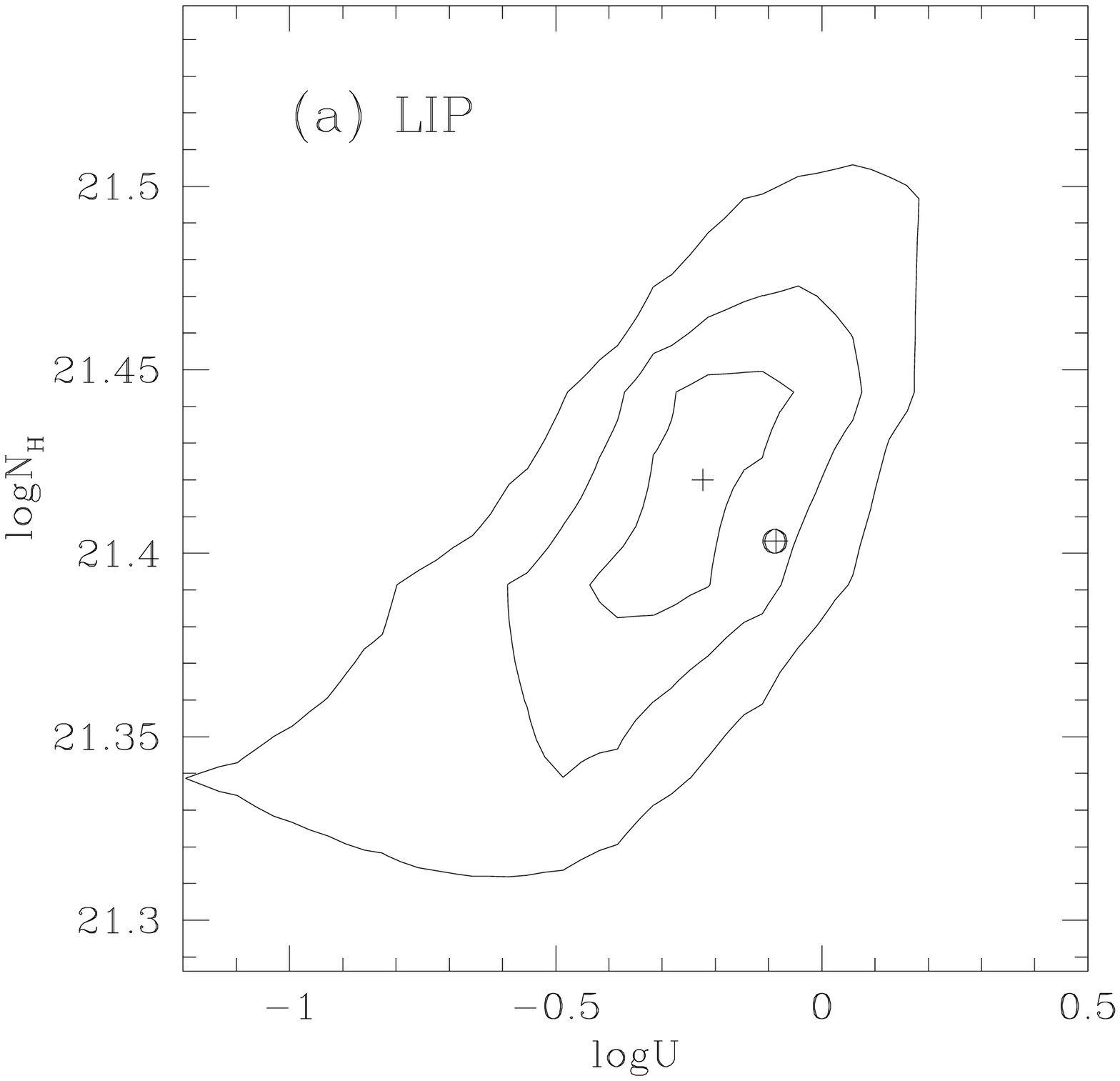}{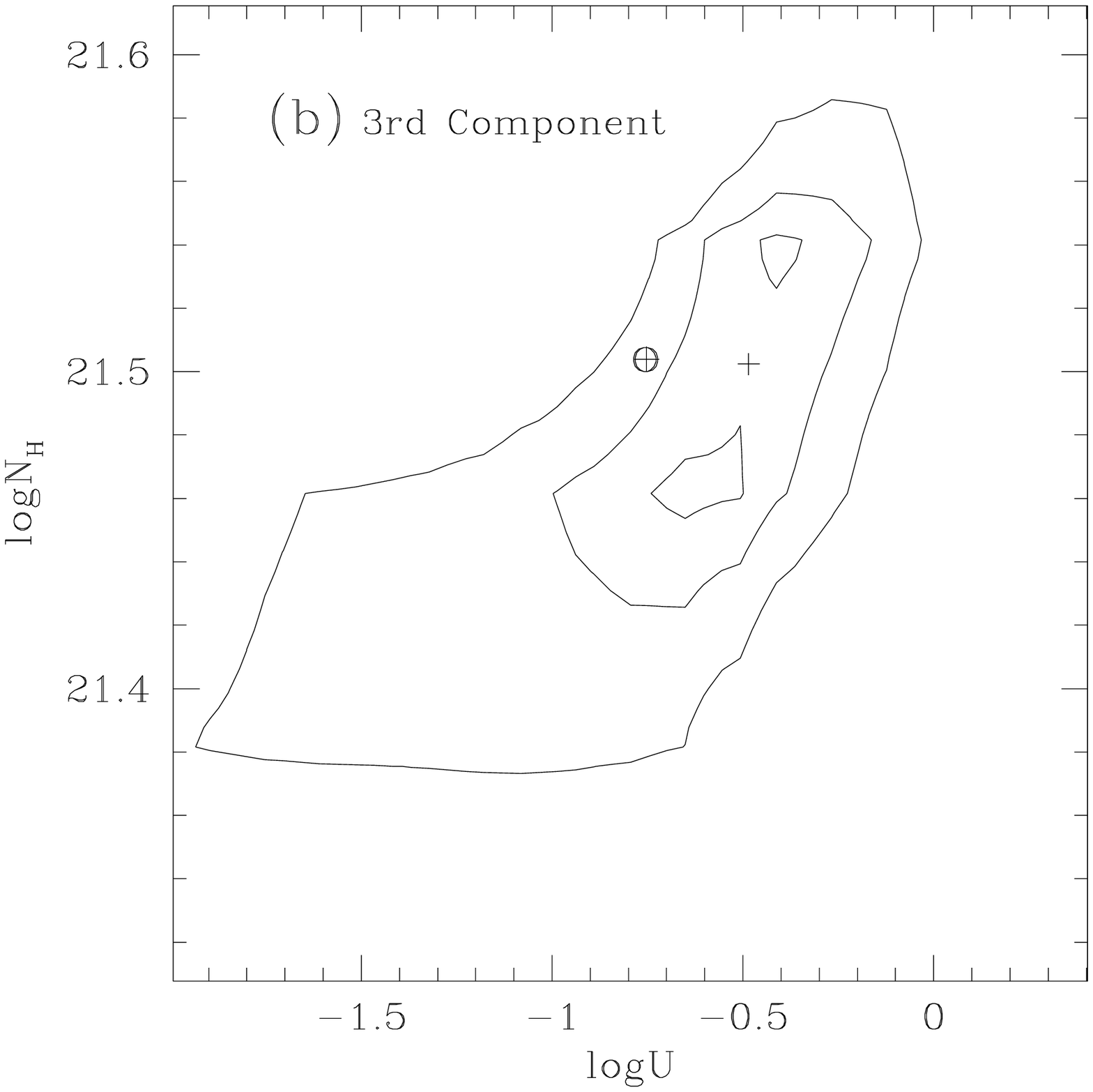}
\caption[f17.eps]{U vs. N$_H$ confidence regions for a model
fitted to simulated data including 3 absorption components: the
HIP and LIP from the {\em Chandra} best fit model plus a third
colder absorption component with $\log$U=-.8 and $\log$N$_H$=21.5
[cm$^{-2}$]. The solid lines correspond to the 1$\sigma$,
2$\sigma$, and 3$\sigma$ levels. In both panels the crossed-circle
point marks the value used to produce the simulated data and the
cross point marks the value obtained in the model. Panel (a):
Confidence region for the LIP. Panel (b): Confidence region for
the third  component. Clearly, a third component would be
detectable in the data, however, the ionization state of such a
component would be difficult to constrain. \label{fake}}
\end{figure}

\noindent (2) The second dataset was simulated using 3 absorption
components: the
HIP and the LIP (Table \ref{2ab}), plus a third cooler
absorber with relatively large column density ($\log U$=-0.8 and
$\log$N$_H$=21.5 [cm$^{-2}$]). First, we modeled these data including only two
absorption components. This approach gives an acceptable fit, but
the ionization parameter and column density of the low ionization gas
are only upper limits, which allows
the presence of the additional cold ionized material.
Then, we fitted the simulated
data with three absorbing components. With this model the values of
the column density and ionization parameter of the LIP can be recovered
within a significance of $\approx 1.5\sigma$, as observed in
Figure \ref{fake}a, and the parameters of the
HIP can also be recovered within a level $\approx 1\sigma$.
The third component is well constrained in N$_H$ (21.4$<$N$_H<$21.6),
but the model
is consistent with the actual values used to simulate this gas at
a significance level of only $\approx 2.5\sigma$, with U poorly
constrained (to a factor 100). This can be observed in Figure
\ref{fake}b, where we present the confidence region for this component.
In fact, our simulations show that the presence of gas 75
times less ionized  than the LIP ($\log U\sim$ -2) would still be
detectable, because the bound-free opacity
increases dramatically with decreasing U. However, the ionization state of
such gas would be basically unconstrained ($\log U<$ -0.5).
Thus, even with low S/N ratio data, we feel
confident that there is no additional absorbing material along the
horizontal branch of the S-curve (i.e. gas out of pressure balance) in
the spectrum of NGC 985, or it would be detected.
Evidently, the high quality data of NGC 3783 is less susceptible
to this kind of bias.

All these tests indicate that, while the pressure-confining
scenario might be appropriate, it requires fragmentation of the
confined absorber, and thus, the idea should be
tested even further.
Any model
of the ionized absorbers around AGN has to
explain several points: (1) Why is there not a significant amount
of gas on the horizontal branch of the S-curve, at temperatures
lower than the one of the LIP? (2) Why does the gas tend to give
preference to special values of the ionization parameter? (3) How do
the observed components relate to the stable branches of the
S-curve, assuming that the gas is indeed in photoionization
equilibrium? (4) What is the relation between the UV and the X-ray
absorbers?
In several objects, the observed X-ray absorbing
components  do not occupy all the
points in the stable branches of the S-curve, and the absorption seems
to be produced from similar charge states. A
simple model with absorbing components located at random
cannot explain these properties. Rather they seem to point to real
properties of the structure of active nuclei.

A perhaps more basic
question that still has to be answered is `where are the X-ray
absorbers located?'
Are they located at large scale distances (parsecs from the
nucleus;  Netzer et al. 2003; Behar et al. 2003) forming part of the
biconical flow observed for the narrow emission line regions
(Kinkhabwala et al. 2002; Behar et al. 2003)? Or, on the contrary,
are they closer (at subparsec distances, e.g. Nicastro et al. 1999;
Netzer et al. 2002; Gabel et
al. 2003; Reeves et al. 2003) and part of a disk wind (Elvis 2000)?

Answers to these questions will help us understand what is the
genesis of these winds and what is their relation to the black
hole and the accretion disk.

\section{ALTERNATIVE SCENARIOS FOR THE IONIZATION MECHANISM \label{col}}

The solution found for the ionized absorber through photoionization
models is satisfactory and carries important implications for the
nuclear environment of AGNs.  Nevertheless, an alternative to
photoionization by the central source deserves to be explored for
the HIP, since
this phase does not respond simply to
continuum changes.

The ring located in NGC 985 at a distance of kiloparsecs from the
nucleus has a large inclination angle, which brings the
possibility that our line of sight to the central source crosses
through collisionally ionized material in this ring. Such collisions
would be produced by an
expanding wave triggered by a pole-on merger (see for instance
Karovska et al. 2002 for Centaurus A). Hence, there is a chance
that the  material producing the absorption observed in
X-rays is collisionally ionized (the HIP
is too hot to be photoionized by the starburst induced by the
merger). If collisions drive the ionization, the temperature of the
HIP is at least
an order of magnitude hotter than our estimate, and thus the
two components  cannot be in pressure equilibrium.
Although this scenario cannot
be excluded, we notice that the change in ionization
structure of the HIP observed in the 3 year period elapsed between
the observations, requires an increase in the electron temperature by
at least a factor of 2 in the HIP (the temperature variation is significant
at a level 2.6$\sigma$), which suggests that this gas is
probably not ionized by collisional mechanisms. Furthermore, if the
absorption of the HIP arises in the ring, then the pressure
balance found between the phases would have to be interpreted as a
pure coincidence. We stress again that pressure balance is also
observed in NGC 3783, a relatively undisturbed galaxy.

\section{SUMMARY}

A new {\em Chandra} observation and an archival BeppoSAX observation
are used to study the nuclear environment of NGC 985.
Absorption features consistent with the presence of ionized
gas in the nuclear environment of this Seyfert 1 galaxy
are detected in both datasets. We have used the code PHASE
(Krongold et al. 2003) to model in a
self-consistent way the ionized absorber, and to further study
possible changes in the opacity of the gas to continuum variability.
Our main results are as follows:.

%\subsection{Analysis of the {\em Chandra} data}

(1) The intrinsic continuum of the source
is well reproduced by a
power law ($\Gamma\approx 1.4-1.6$) and a thermal component (kT=0.1 keV). This
continuum was attenuated by an equivalent hydrogen column density
of $3.0\times10^{20}$ cm$^{-2}$ to account for Galactic
absorption.

(2) The data requires two absorption components different by a factor
of 29 in ionization parameter, and by a factor of 3 in
H column density. The hotter phase
is clearly identified through broad absorption features produced by
several blends of Fe L-shell absorption, and the cooler component is
detected due to the presence of a deep Fe M-shell UTA. However, the
signal to noise ratio of the {\em Chandra} data does not allow the
identification of significant ($\sigma > 2$) narrow (unblended)
absorption features. This results
in a poor determination of the outflow velocities of the components ($581\pm
206$ km s$^{-1}$ for the high ionization phase and $197\pm
184$ km s$^{-1}$ for the low ionization one).

% high ionization phase (U=21.84) is
%$\approx 29$ times larger than that of the low ionization phase
%(U=0.76). The equivalent H column densities were estimated as
%N$_H=6.5\times 10^{21}$ cm$^{-2}$  for the hotter phase and
%N$_H=2.5\times 10^{21}$ cm$^{-2}$ for the cooler one.

(3) Despite the limited signal to noise ratio of the {\em Chandra}
data, the ionization state of both components is well constrained
thanks to several features of 
Fe L-shell absorption, but mainly because of the Fe UTA. The lack of accurate
dielectronic recombination rates of low charge states of Fe (Netzer et
al. 2003) introduces
an uncertainty
by a factor less than 2 in the ionization parameter of the 
phases; however,  our results are robust and highly insensitive to
this source of error.

(4) The outflow velocity of both components is consistent with most of
the absorption components found in the UV spectrum of this object. In
particular, both components are consistent with the UV component with
outflow velocity of 404 km s$^{-1}$, which is hot enough to absorb in
the X-ray regime (Arav 2002). In addition, the low ionization
component in the X-rays produces significant amounts of O {\sc vi}, N
{\sc v}, and C {\sc iv}, which strongly suggests that the absorption
features in both bands are the manifestation of a single
wind.

(5) The data set tight limits on the presence of additional cold
absorption components, showing that gas with temperature $<10^6$ K
clumps around the two
absorbers detected. On the other hand, the presence of a third
component, hotter than the other two (T$>10^6$ K), and producing Fe
K-shell absorption,
is hinted by the data. However, data with better signal to noise ratio
is needed to constrain the properties of such component.

(6) The gas detected in the spectrum of NGC 985 has strikingly
similar characteristics to the gas found in NGC 3783 and IRAS
13349+2438: (1) absorption produced by cold gas (Fe {\sc
vii-xiii}), (2) absorption by hot gas, dominated by Fe {\sc
xvii-xxii}, and (3) no significant absorption consistent with Fe
{\sc xiv-xvi}. Then a continuous range of ionization parameters
is disfavored (at least in these objects). These similar
characteristics further suggest that this is an intrinsic property
of the absorbers, probably related to the unstable branches of the
thermal equilibrium curve.

(7) The X-ray luminosity drops by a factor 2.3 between the
BeppoSAX and {\em Chandra} observations (separated by $\sim 3$
years). We detect (model dependent) variability in the opacity of
the absorbers. While the colder component is consistent with a
simple picture of photoionization equilibrium, the ionization
state of the hotter component seems to increase while the flux
drops.

(8) The most striking result in our analysis is that the two absorbing
phases are consistent (to within the errors) with pressure
equilibrium. This result is
obtained during both
the {\em Chandra} and the BeppoSAX
observations. Based on this, we speculate that the absorption arises from
a multi-phase wind. Such 
a scenario can explain the change in
opacity of both components during the observations, but requires that
a third hotter component is pressure-confining the two phases.
Thus, our analysis strongly points to a 3-phase medium, just like the
wind suggested in NGC 3783.

(9) An important implication of our analysis is that, if the
pressure-confining scenario is real, then fragmentation of the confined
phases in a large number of clouds is required.

The analysis presented in this paper provides further evidence that warm
absorbers in AGN are simple systems that can be relatively well
understood. The high resolution 
spectroscopic capabilities of the new
generation of X-ray observatories have revealed warm
absorbers as strong winds, now recognized as an important feature
of AGN structure. Here, we have used high and low resolution data
to constrain and characterize the physical conditions and structure of these
winds. Measuring the
response in the
opacity to variability of the ionizing flux, we suggest that the
nature of the ionized
absorber is that of a multi-phase wind, with one phase confining the
other two.
The implications of this result are extremely important to our
understanding of quasar energy budgets, as well as to our
understanding of the
interaction between these winds 
and the interstellar medium of the
host galaxy. However, our analysis is based on a comparison between
high and low resolution data. Only a direct comparison
between high
resolution data will allow further testing of the
scenario presented here, leading to a deeper
understanding of the nature of AGN and quasars.

\acknowledgements We thank the anonymous referee for valuable comments
that helped improve our work. This research has been partly supported by NASA
Contract NAS8-39073 (Chandra X-ray Center), NASA grant NAS G02-3122A,
and Chandra General Observer Program TM3-4006A.
This research has made use of the NASA/IPAC Extragalactic Database
(NED) which is operated by the Jet Propulsion Laboratory, California
Institute of Technology, under contract with the National Aeronautics
and Space Administration.

\clearpage

\begin{deluxetable}{lccc}
\tablecolumns{4} \tablewidth{0pc} \tablecaption{Observation Log of NGC
985\label{log}}

\tablehead{\colhead{Observatory} & \colhead{Sequence Number} &
\colhead{UT Start} & Exp Time (ks)$^a$ }
\startdata
BeppoSAX &  50862001   & 1999 AUG 20 06:38  & 95.0  \\
{\em Chandra} & 700449 & 2002 JUN 24 19:33  & 77.7  \\
\enddata
\tablenotetext{a}{Sum of good time intervals.}
\end{deluxetable}

\clearpage

\onecolumn

\begin{deluxetable}{lccc}
\tablecolumns{4} \tablewidth{0pc} \tablecaption{Continuum
Parameters \label{tabcont}}

\tablehead{\colhead{Parameter} & \multicolumn{2}{c}{Chandra Data} &
\colhead{BeppoSAX Data} \\
& \colhead{0.7-8 keV$^a$} & \colhead{3-8 keV} & \colhead{0.1-110 keV$^a$} }
\startdata
$\Gamma^b$ &  1.60$\pm 0.03$ &  1.62$\pm 0.02$       & 1.40$\pm .04$ \\
Norm$^c$   &  1.22$\pm 0.04$ &  1.22$\pm 0.04$      &2.47$\pm 0.04$ \\
BB kT$^d$  & 0.10$\pm 0.01$   & \nodata      & 0.10$\pm 0.02$ \\
Norm$^e$   & 3.7$\pm 0.8$  & \nodata         &8.8$\pm 1.2$ \\
E$_c$ (keV)$^f$  & \nodata & \nodata         &$67\pm 24$ \\
Flux(.1-2keV)$^g$ & 0.63$\pm 0.12 $ & \nodata          &  1.34$\pm 0.08$\\
Flux(2-10keV)$^g$ & 0.54$\pm 0.09$  & \nodata          &1.49$\pm 0.12$\\
R$^h$            & \nodata          & \nodata           &  0.0$+0.35$\\
$\chi^2/$d.o.f. &  179.0/174  &  46.7/49    &  116.4/110    \\
\enddata
\tablenotetext{a}{Continuum fits carried out with the
self-consistent inclusion of
an ionized absorber. See Tables \ref{2ab} and \ref{tab_abs_sax}.}
\tablenotetext{b}{Power Law Photon Index} \tablenotetext{c}{Power
law normalization in 10$^{-3}$ photons keV$^{-1}$ cm$^{-2}$ s$^{-1}$ at 1
keV} \tablenotetext{d}{Blackbody Temperature in keV}
\tablenotetext{e}{Blackbody Normalization in 10$^{-5}$ $L_{39}/D_{10}^2$,
where $L_{39}$ is the source luminosity in units of $10^{39}$ erg
s$^{-1}$  and $D_{10}$ is the distance to the source in units of
10 kpc} \tablenotetext{f}{High Energy Cutoff as inferred
from the SAX data} \tablenotetext{g}{In units of $10^{-11}$ ergs cm$^{-2}$ s$^{-1}$}
\tablenotetext{h}{Reflection component factor
(R=$\Delta \Omega$/2$\pi$, with $\Delta \Omega$ the solid angle
subtended by the reflector).}
\end{deluxetable}

\clearpage

\begin{deluxetable}{lcc}
\tablecolumns{3} \tablewidth{0pc} \tablecaption{Two Phase Absorber
Parameters for {\em Chandra} Data \label{2ab}}
\tablehead{\colhead{Parameter} & \colhead{High-Ionization} &
\colhead{Low-Ionization} \\ & \colhead{HIP} &
\colhead{LIP}} \startdata
Log U$^a$                     &  1.34$\pm0.10$  & -0.12$^{+0.30}_{-0.09}$   \\
Log N$_{H}$ (cm$^{-2}$)$^a$   &  21.81$\pm.22$ & 21.39$\pm0.19$ \\
V$_{Turb}$ (km s$^{-1}$)$^b$      &  350           & 350            \\
V$_{Out}$ (km s$^{-1}$)$^a$   & 581$\pm 206$    & 197$\pm 184$     \\
$[$Log T (K)$]$$^c$                    & 5.79$\pm0.11$ & 4.47$\pm 0.04$ \\
Log U/T ($\propto$1/P$^d$)           &-4.45 $\pm0.21$&
-4.59$^{+0.30}_{-0.13}$ \\
Log U$_x^e$                     &-1.09$\pm0.10$  & -2.55$\pm0.09$   \\
\enddata
\tablenotetext{a}{Free parameters of the model.}
\tablenotetext{b}{Fixed parameter.}
\tablenotetext{c}{Derived from the column density and ionization
parameter, assuming photoionization equilibrium.}
\tablenotetext{d}{The gas pressure P$\propto$n$_e$T. Assuming that
both phases lie at the same distance from the central source
n$_e\propto$1/U, and P$\propto$T/U.}
\tablenotetext{e}{Ionization parameter defined in the 0.1-10 keV range
(Netzer 1996).}
\end{deluxetable}

\clearpage

\begin{deluxetable}{lcc}
\tablecolumns{3} \tablewidth{0pc} \tablecaption{Two Phase Absorber
Parameters for BeppoSAX Data \label{tab_abs_sax}}
\tablehead{\colhead{Parameter} & \colhead{High-Ionization} &
\colhead{Low-Ionization} \\ & \colhead{HIP} &
\colhead{LIP} } \startdata
Log U$^a$                     &  0.83$\pm0.13$  &-0.22$^{+0.30}_{-0.10}$ \\
Log N$_{H}$ (cm$^{-2}$)$^b$   &  21.81        & 21.39 \\
V$_{Turb}$ (km s$^{-1}$)$^b$      &  350           & 350            \\
V$_{Out}$ (km s$^{-1}$)$^b$   & 581      & 197        \\
$[$Log T (K)$]$$^c$                    & 5.48$\pm0.12$ & 4.56$\pm 0.05$ \\
Log U/T ($\propto$1/P$^d$)           &-4.65 $\pm0.25$&
-4.78$^{+0.30}_{-0.15}$ \\
Log U$_x^e$                     &-1.25$\pm0.13$  & -2.30$\pm0.10$   \\
\enddata
\tablenotetext{a}{Free parameter of the model.}
\tablenotetext{b}{Fixed parameters, assuming N$_H$, FWHM, and
V$_{Out}$ values from the best fitting {\em Chandra} model.}
\tablenotetext{c}{Derived
from the column density and ionization parameter, assuming
photoionization equilibrium.} \tablenotetext{d}{The gas pressure
P$\propto$n$_e$T. Assuming that both phases lie at the same
distance from the central source n$_e\propto$1/U, and
P$\propto$T/U.} \tablenotetext{d}{Ionization parameter defined in
the 0.1-10 keV range (Netzer 1996).}
\end{deluxetable}

\clearpage

\begin{deluxetable}{lcclcc}
\tablecolumns{6} \tablewidth{0pc} \tablecaption{Ionic Column Densities$^a$
predicted by the Models\label{ionic}}

\tablehead{\multicolumn{2}{c}{HIP} & \colhead{} & \multicolumn{2}{c}{LIP} \\
\colhead{Ion} & \colhead{\em Chandra} & \colhead{BeppoSAX} &
\colhead{Ion} & {\em Chandra} & \colhead{BeppoSAX} }
\startdata
C {\sc vi} & 16.53 & 16.88	& C {\sc v} & 17.31  & 17.01  \\
C {\sc vii} & 18.34 & 18.36	& C {\sc vi} & 17.64 & 17.68  \\
N {\sc vii} & 16.46 & 16.80	& N {\sc vi} & 17.03 & 16.94  \\
N {\sc viii} & 17.76 & 17.73	& N {\sc vii} & 16.95& 16.97   \\
O {\sc vii} & 16.18 & 16.94	& O {\sc vii} & 18.03& 18.07  \\
O {\sc viii} & 17.75 & 18.09	& O {\sc viii} & 17.48& 17.69 \\
O {\sc ix} & 18.61 & 18.54	& Ne {\sc vi} & 17.16 & 16.97 \\
Ne {\sc ix} & 16.52 & 17.09	& Ne {\sc vii} & 16.70& 16.77 \\
Ne {\sc x} & 17.38 & 17.54	& Ne {\sc viii} & 16.44& 16.75 \\
Ne {\sc xi} & 17.66 & 17.45	& Ne {\sc ix} & 16.20 & 16.78  \\
Mg {\sc xi} & 16.55 & 16.99	& Mg {\sc viii} & 16.51& 16.21 \\
Mg {\sc xii} & 17.10 & 17.02 	& Mg {\sc ix} & 16.54 & 16.57 \\
Mg {\sc xiii} & 16.91 &	16.56	& Mg {\sc x} & 16.10 & 16.45 \\
Si {\sc xiii} & 16.96 & 17.09	& Si {\sc viii} & 16.29& 15.76 \\
Si {\sc xiv} & 16.98 & 16.79	& Si {\sc ix} & 16.52 & 16.34 \\
Si {\sc xv} & 16.47 & 16.03	& Si {\sc x} & 16.37 & 16.53 \\
Fe {\sc xvii} & 16.16 &	16.60 	& Fe {\sc ix} & 16.01& 15.28 \\
Fe {\sc xviii} & 16.69 & 16.83	& Fe {\sc x} & 16.24 & 15.77 \\
Fe {\sc xix} & 16.81 &	16.70	& Fe {\sc xi} & 16.34& 16.11 \\	
Fe {\sc xx} & 16.72 &	16.36	& Fe {\sc xii} & 16.19& 16.33\\	
Fe {\sc xxi} & 16.28 &	15.68	& Fe {\sc xiii} & 15.92& 16.28 \\	
\enddata
\tablenotetext{a}{$\log$ of column density in cm$^{-2}$}
\end{deluxetable}

\clearpage

\begin{deluxetable}{lcc}
\tablecolumns{3} \tablewidth{0pc} \tablecaption{Column densities$^a$ of
low ionization LIP ions from the {\em Chandra} best fit model. \label{uvtab}}
\tablehead{\colhead{Ion} & \colhead{SED1$^b$} &\colhead{SED2$^c$}}
\startdata
O {\sc vi} &   17.23 & 17.13   \\
N {\sc v}  &   15.67 & 15.39    \\
C {\sc iv} &   15.82 & 15.23    \\
H {\sc i}  &   15.37 & 15.33    \\
\enddata
\tablenotetext{a}{$\log$ of column density in cm$^{-2}$}
\tablenotetext{b}{SED presented in Fig. \ref{fsed}a}
\tablenotetext{a}{SED presented in Fig. \ref{fsed}b}
\end{deluxetable}

\clearpage

\clearpage

\clearpage

\clearpage

\clearpage

\clearpage

\clearpage

\clearpage

\clearpage

\clearpage

\clearpage

\clearpage

\clearpage

\clearpage

\clearpage

\clearpage

\clearpage

\clearpage

\twocolumn

\end{document}